\documentclass[12pt,pra,aps]{revtex4}
\usepackage{graphicx}
\usepackage{dsfont,psfrag}
\usepackage{amsgen,amsfonts,amssymb,amsbsy,latexsym,amsthm,amscd}
\usepackage{amsmath}
\usepackage{times}
\usepackage{epsfig}
\usepackage{textcomp}
\usepackage{rotating}
\usepackage{here}
\usepackage{bm}
\usepackage[latin1]{inputenc}

\newcommand{\kt}{{\tilde{k}}}
\newcommand{\kc}{k_c}

\newcommand{\omt}{{\tilde{\omega}}}
\newcommand{\omc}{\omega_c}
\newcommand{\omp}{\omega_{+}}
\newcommand{\omm}{\omega_-}
\newcommand{\omk}{\omega_k}
\newcommand{\Tr}{\text{Tr}}

\newcommand{\xit}{{\tilde{\xi}}}
\newcommand{\xic}{\xi_c}
\newcommand{\xip}{\xi_{+}}

\begin{document}

\title{Exploring the growth of correlations in a quasi one-dimensional trapped
  Bose gas} 

\author{M. Eckart, R. Walser and W.P. Schleich} 
\address{Institute of Quantum Physics, Ulm University, D-89069
  Ulm, Germany} \date{\today}
\email{Michael.Eckart@uni-ulm.de}
\begin{abstract}
  Phase correlations, density fluctuations and three-body loss rates are
  relevant for many experiments in quasi one-dimensional geometries. Extended
  mean-field theory is used to evaluate correlation functions up to third
  order for a quasi one-dimensional trapped Bose gas at zero and finite
  temperature. At zero temperature and in the homogeneous limit, we also 
  study the transition from the weakly correlated Gross-Pitaevskii regime to 
  the strongly correlated Tonks-Girardeau regime analytically. We compare our 
  results with the exact Lieb-Liniger solution for the homogeneous case and 
  find good agreement up to the cross-over regime.
\end{abstract}

\maketitle

\pagebreak

\tableofcontents

\pagebreak

\section{Introduction}

Low dimensional physics has always been in the focus of interest as
reduced dimensionality goes hand in hand with an increase of quantum
fluctuations. Thus, by reducing the volume of accessible phase space, the
effect of quantum fluctuations of the remaining degrees of freedom must be
enhanced in order to comply with the fundamental Heisenberg uncertainty
principle.

Currently, this fact has stimulated many fascinating experiments in the
context of ultracold gases \cite{Goerlitz01, Esslinger03, Tolra04, 
Ertmer03,Weiss05, Raizen2005, Druten07, schmiedmayer}, which explore
various aspects of the geometric transitions.  Serendipitously, also most
exactly solvable models of field theory are one-dimensional
\cite{MathisLiebBuch} and rest on the celebrated Bethe-ansatz invented in the
1930ies. In the context of atomic Bose gases, today most prominent are the
spatially homogeneous models of hard- and soft-core bosons on a string of
M.~Girardeau \cite{gir60} as well as E.~Lieb and W.~Liniger
\cite{Lieb63I,Yang69}.  One of the many interesting questions which can be
explored, is the cross-over from the weakly correlated Gross-Pitaevskii regime
($\gamma \ll 1$) to the strongly correlated Tonks-Girardeau regime
\cite{Das2002, Paredes2004} ($\gamma \gg 1$). Thereby, one commonly uses the
Lieb-Liniger (LL) parameter $\gamma$, cf. (\ref{LLparam}), to measure the
relative strength of kinetic to repulsive self-energy in the dilute Bose gas.

Today, we also have alternative tools available: First, there are the exact
few-body calculations, i.\thinspace{}e.~multi-channel time-dependent
Hartree-Fock (MCTHF) or configuration interaction (CI) methods, which
originate from atomic, molecular and nuclear physics. While originally
designed for fermionic energy structure calculation, they are nowadays also
applied to few-boson systems ($\approx$~10-100 particles) in arbitrary trap
geometries \cite{esry97,ofir_alon04,zollner2006a,cederbaum06}.  Second, there
is now the possibility to prepare atomic gases in optical lattices and this
opens up the rich methodology of the Bose-Hubbard model and density matrix
renormalization group methods
\cite{whitedmrg,jaksch98,vanOosten2001a,Muramatsu05,Zwerger04,Ingvarson_half_filling}.
Third, there are stochastic multi-mode trajectory simulations
\cite{drummond03} that also successfully address the same questions.

Irrespective of the choice of method, all need to predict experimentally
accessible observables in terms of correlation functions, spatial averages or
Fourier transforms, thereof.  Most relevant are obviously the lowest order
moments of the bosonic field operator $\hat{a}_x$, which is the single
particle density $n_x$ at position $x$ and the conjugate phase quadrature
correlation function $g^{(1)}_{x,y}$. The fluctuations about the mean density
are measured with the second order density-density correlation function
$g^{(2)}_{x,y}$
\begin{equation}
  \label{g1g2}
  n_x=\langle \hat{a}_x^{\dagger}\hat{a}_x^{\phantom{\dagger}}\rangle,
  \qquad
  g^{(1)}_{x,y} = \frac{\langle \hat{a}_y^{\dagger}
    \hat{a}_x^{\phantom{\dagger}} \rangle}{\sqrt{n_x n_y}},
  \qquad
  g^{(2)}_{x,y}
  = \frac{\langle \hat{a}_x^{\dagger}\hat{a}_y^{\dagger}
    \hat{a}_y^{\phantom{\dagger}}\hat{a}_x^{\phantom{\dagger}} \rangle}{n_x
    n_y}. 
\end{equation}
In here, $\langle \ldots\rangle=\Tr\{ \ldots \boldsymbol{\rho}\}$
denotes an average over the state of the system described by the many-body
density operator $\boldsymbol{\rho}$. Such second order correlation functions
have been measured experimentally
\cite{Ertmer03,Weiss05,shimitsu96,Kasevich,CohenTannoudji97,westbrook07},
while the third order density-density-density correlation
\begin{equation}
  \label{g3}
  g^{(3)}_{x,y,z} = \frac{\langle \hat{a}_x^{\dagger}\hat{a}_y^{\dagger} 
    \hat{a}_z^{\dagger}
    \hat{a}_z^{\phantom{\dagger}} \hat{a}_y^{\phantom{\dagger}} 
    \hat{a}_x^{\phantom{\dagger}} \rangle}{n_x n_y n_z},
\end{equation}
became observable only recently \cite{Tolra04,cornell597} via the three-body
recombination rate \cite{Shlyap85}.  

Theoretically, much attention has been
directed towards second order correlation functions \cite{Petrov00,
  Olshanii03, ShlyapGang03, Shlyap03, walser04, Bog04,Shlyap05, Giorgini05},
while less is known about the third order correlation function. This situation
has been rectified recently in \cite{Cheianov2006a, Cheianov2006b} where the
diagonal behaviour of this correlation function was calculated in the framework
of Lieb-Liniger theory.  This is where extended mean-field theory is useful,
because we can calculate arbitrary orders of the correlation function and will
present calculations of the diagonal and off-diagonal behaviour of the third
order correlation function at zero as well as finite temperature. However, the
extended mean-field (EMF) approach is restricted to values of the correlation
parameter $\gamma \leq 1$, because any mean-field theory is known to fail in
the strongly correlated regime.

This paper is organized as follows: In section~\ref{Lieb}, we briefly review
the central ideas of Lieb-Liniger theory \cite{Lieb63I}. This celebrated
solution of the one-dimensional homogeneous Bose gas is an ideal benchmark for
the extended mean-field theory
\cite{walser04,Walser99,holland301,kokkelmans501}, whose basic concepts are
summarized in section~\ref{theory}. In section~\ref{secstathfb}, we specialize
the kinetic equations to a quasi one-dimensional homogeneous situation at zero
temperature for which analytical solutions can be found and compare
correlation functions with the LL-predictions.  The extension to
inhomogeneous, harmonically trapped systems at finite temperatures is studied
numerically in section~\ref{zerofinite}.

\pagebreak

\section{Lieb-Liniger theory for bosons in one dimension}
\label{Lieb}
Lieb-Liniger theory based on the Bethe ansatz \cite{MathisLiebBuch} describes
a one-dimensional homogeneous gas of $N$ bosons on a ring of length $L$. It is
one of very few exactly solvable problems in many-body physics and provides a
solution for every value of the correlation parameter $\gamma$.  Even in
inhomogeneous trapped systems this is very useful, if we can make the local
density approximation.  In the language of second quantization, the starting
point for Lieb-Liniger theory is the following Hamiltonian
\begin{equation} 
  \hat{H} = \int_0^L {\rm d}{x}\, 
    \hat{a}^{\dagger}_{{x}}\left(
      -\frac{\hbar^2}{2m}
      \frac{\partial^2}{\partial x^2}
      + \frac{g}{2}
      \hat{a}^{\dagger}_{{x}}\hat{a}^{\phantom{\dagger}}_{{x}}
  \right)\hat{a}^{\phantom{\dagger}}_{{x}},
\end{equation}
where $m$ denotes the mass of a boson and the creation and annihilation
operators satisfy the usual bosonic commutation relation.
With the help of the Hellmann-Feynman theorem \cite{Feynman39}, one can obtain
the diagonal part ($x=y=0$) of the translation invariant second order
correlation function $g^{(2)}_{x,y}=g^{(2)}_{{ LL}}$, introduced in
(\ref{g1g2}), as
\begin{equation} 
  \label{g2_1}
  \frac{L}{2}{g^{(2)}_{{ LL}}} n^2=
  \frac{d E_0}{d g} = \langle \Psi_0
  |\frac{d \hat{H}}{d g}|\Psi_0\rangle
\end{equation}
by differentiating the ground state energy $E_0$ with respect to the coupling
constant $g$. Here, $| \Psi_0 \rangle$ represents the ground state and
$n=N/L$ denotes the linear particle density. It was shown by Lieb and Liniger
\cite{Lieb63I} that the ground state energy only depends on the dimensionless
correlation parameter $\gamma$. It is basically the ratio of the repulsive mean-field
energy $g n$ to the kinetic energy $\hbar^2/2m d^2$ at an average distance
$d=1/n$. Another length scale of the problem is the healing length $\xi$,
which equates the kinetic energy of a wave function at scale $\xi$ to the mean-field energy
\begin{equation}
  \label{LLparam}
  \gamma = \frac{m g}{\hbar^2 n}, \qquad \xi=\frac{\hbar}{\sqrt{2 mn g}} \; .
\end{equation}
 We call bosons weakly
correlated for $\gamma\ll 1$ (Gross-Pitaevskii regime) and strongly correlated
for $\gamma \gg 1$ (Tonks-Girardeau regime). 

In terms of this parameter, the
ground state energy and second order correlation function
\begin{equation}
  \label{e_gamma}
  E_0 =N \frac{\hbar^2 n^2}{2m} e(\gamma),\qquad 
  g^{(2)}_{{ LL}}=e'(\gamma),
\end{equation}
are given in terms of the solutions of the Lieb-Liniger equations
\begin{eqnarray}
  e(\gamma)&=& \frac{\gamma^3}{\lambda^3(\gamma)}\int_{-1}^1 {\rm d}x\,
  h(x,\gamma)x^2 \\
  \label{LLeqs}    
  h(x,\gamma)&=&\frac{1}{2\pi}+\frac{1}{\pi}\int_{-1}^1 {\rm d}y\,
  \frac{\lambda(\gamma) h(y,\gamma)}{\lambda^2(\gamma) + (y-x)^2},\quad
  \lambda(\gamma) = \gamma \int_{-1}^1 {\rm d}x\,
  h(x,\gamma).
\end{eqnarray}

\begin{figure}
  \includegraphics[width=0.8\columnwidth]{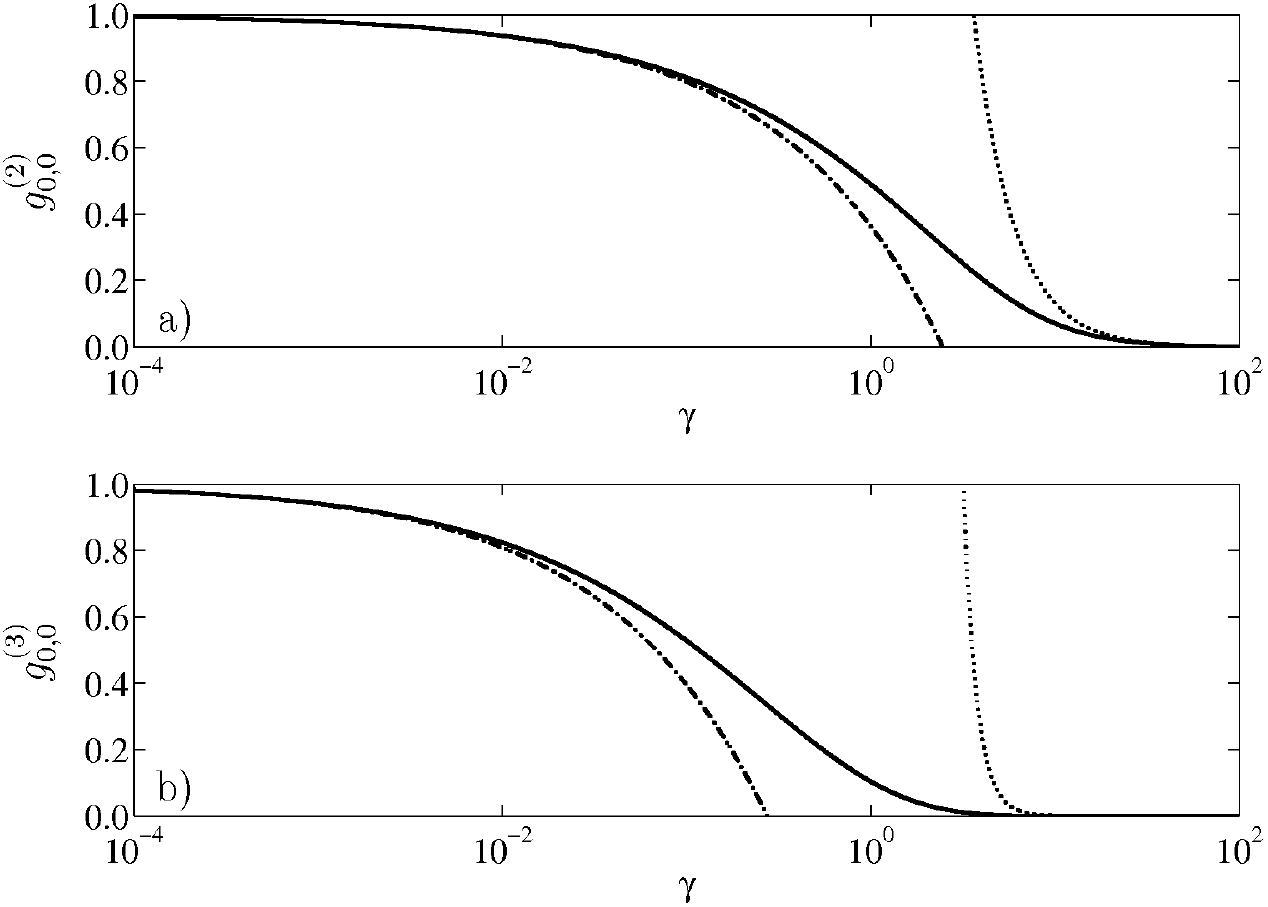}
  \caption{The Lieb-Liniger correlation function
    versus the correlation parameter $\gamma$.  In subplot a), we depict the
    second order correlator $g^{(2)}_{ LL}$ (solid line), the
    Gross-Pitaevskii approximation $g^{(2)}_{{ LL,GP}}$ (dashed dotted
    line) and the Tonks-Girardeau approximation $g^{(2)}_{{ LL,TG}}$ (dotted line), while
    subplot b) shows the third order correlator $g^{(3)}_{{ LL}}$ (solid
    line) and the approximations $g^{(3)}_{{LL,GP}}$ (dashed dotted line) and
    $g^{(3)}_{ {LL,TG}}$ (dotted line).}
\label{Lieb_pic_g2}
\end{figure}

In the weakly correlated Gross-Pitaevskii limit $\gamma \to 0$, as well as in
the strongly correlated Tonks-Girardeau regime $\gamma \to \infty$, one
obtains for the correlation function \cite{Shlyap03}
\begin{equation}
  g^{(2)}_{ {LL,GP}}  =   
  1- \frac{2\sqrt{\gamma}}{\pi}, \mbox{ for }\gamma \ll 1,\qquad
  g^{(2)}_{ {LL,TG}} = 
  \frac{4\pi^2}{3\gamma^2}, \mbox{ for }\gamma \gg 1.
\end{equation} 
A comparison of these approximations with the exact solution is presented in
figure \ref{Lieb_pic_g2}. The validity of these results has recently been tested
experimentally over a wide range of the correlation parameter \cite{Weiss05}
and will be used to probe the extended mean-field approach presented in the
next section.

\pagebreak

It is also possible to derive an exact result for the diagonal part of the
third order correlation function within the framework of Lieb-Liniger theory,
however the task is considerably more difficult. The exact result is derived
in \cite{Cheianov2006b} by introducing a new function $\tilde{e}(\gamma)$
which has the form
\begin{eqnarray}
  \label{e4_gamma}
  \tilde{e}(\gamma) &=& \frac{\gamma^5}{\lambda^5(\gamma)}
  \int_{-1}^1{\rm d}x\, h(x,\gamma)x^4
\end{eqnarray}
and with the help of this function one obtains
\begin{eqnarray}
  \label{Chei_g3}
  g^{(3)}_{{ LL}} &=& \frac{3\tilde{e}'(\gamma)
    -4 e(\gamma)-6 e(\gamma)e'(\gamma)}{2\gamma} +
  \left(1+\frac{\gamma}{2}\right)e'(\gamma) 
  + \frac{9 {e}^2(\gamma)-5 \tilde{e}(\gamma)}{\gamma^2}.
\end{eqnarray}

\begin{figure}
\includegraphics[width=0.8\columnwidth]{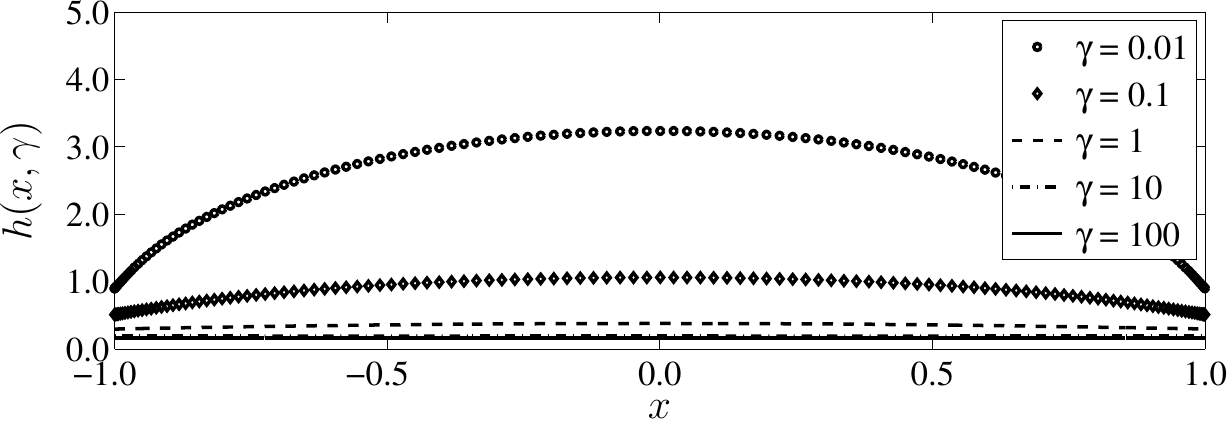}
\caption{
    The auxiliary Lieb-Liniger function $h(x,\gamma)$, as a function of the
    variable $x$ for five different values of the correlation parameter
    $\gamma$.}
\label{Lieb_g_x}
\end{figure}

A comparison of the exact result for the third order correlation function with
the approximations in the Gross-Pitaevskii and the Tonks-Girardeau regime
\cite{Shlyap03} is presented in figure \ref{Lieb_pic_g2}
\begin{equation}
  g^{(3)}_{{LL,GP}} =
  1- \frac{6\sqrt{\gamma}}{\pi}, \mbox{  for }\gamma \ll 1, \qquad
  g^{(3)}_{{LL,TG}} =  \frac{16\pi^6}{15\gamma^6}, 
  \mbox{ for } \gamma \gg 1.
\end{equation} 
The auxiliary function $h(x,\gamma)$ of (\ref{LLeqs}) is depicted for
various values of the correlation parameter in figure \ref{Lieb_g_x}.

\pagebreak

\section{Extended mean-field theory for bosons in one dimension}
\label{theory}

\subsection{Time-dependent Hartree-Fock-Bogoliubov equations}
\label{TDHFB_equations}
The evolution of a weakly interacting dilute gas of bosons in three dimensions
can be described by a Hamiltonian
\begin{eqnarray}
  \label{3DHamiltonian}
  \hat{H}  &=&  \int {\rm d}^6  xy\,
  \hat{a}^{\dagger}_{\mathbf{x}}\left[
    \mathcal{H}^{\phantom{\dagger}}_{\mathbf{xy}}
    +\frac{1}{2}  V_{bin}(\mathbf{x}-\mathbf{y})
    \hat{a}^{\dagger}_{\mathbf{y}}
    \hat{a}^{\phantom{\dagger}}_{\mathbf{y}} \right]
  \hat{a}^{\phantom{\dagger}}_{\mathbf{x}},\\
  \label{eqm_end}
  \mathcal{H}_{\mathbf{xy}}  &=& 
  \langle\mathbf{x}|\frac{\mathbf{p}^2}{2m}+
  V_{ext}(\mathbf{x})|\mathbf{y}\rangle,
  \quad
  V_{ext}(\mathbf{x}) = \frac{1}{2}m \omega^2x^2 + 
  \frac{1}{2}m \omega_{\perp}^2(y^2 + z^2),
\end{eqnarray}
where $\mathcal{H}$ is the single-particle energy in an external potential
$V_{ext}$ and $V_{bin}$ is the two-particle potential.  As we are interested
in the quasi one-dimensional limit, we will consider a cigar shaped trapping
configuration (angular frequencies $\omega$ and $\omega_\perp$) 
with a large aspect ratio $\beta$. The energy and length scales will be
set by the transverse oscillator   
\begin{equation}
   \beta =\omega_{\perp}/\omega\gg 1, 
\quad a_\perp = \sqrt{\hbar/m\omega_\perp},\quad
 \varepsilon_\perp=\hbar\omega_\perp.
\end{equation}

Conceptually, it is straight forward in the extended mean-field theory to use
real, finite range binary interaction potentials and obtain proper two-body T
matrices including many-body corrections.  However, for convenience, we will
use the pseudo-potential approximation in here
\begin{equation}
  V_{bin}(\mathbf{x}-\mathbf{y}) = \frac{4\pi \hbar^2 a_s}{m} 
\delta_{\mathbf{x-y}},
\end{equation}
where $a_s$ denotes the s-wave scattering length.  In order to compactify the
notation we will interchangeably use a subscript notation also for continuous
functions.

Extended mean-field theory uses a reduced state description based on a set of
master variables $\{i\in I|\gamma_i\}$. Basically, this implies the existence
of a well separated hierarchy of time, energy and length scales
\cite{peletminskii,zubarev1} and leads to a rapid attenuation of
correlation functions. Mathematically speaking, it allows for a selfconsistent
expansion of the full many-body density matrix $\boldsymbol{\rho}$ in terms of
a perturbation series of simple many-body density matrices
$\boldsymbol{\sigma}^{(i)}$, which depend parametrically on the
master variables 
\begin{equation}
  \label{pertseries}
  \boldsymbol{\rho}=\boldsymbol{\sigma}^{(0)}_{\{\gamma\}}
  +\boldsymbol{\sigma}^{(1)}_{\{\gamma\}}+\mathcal{O}(V_{bin}^2). 
\end{equation}
This non-perturbative series in terms of the interaction potential $V_{bin}$
has been introduced first by Chapman and Enskog in the context of kinetic
theory of gases \cite{chapman}.  In addition to the simple series expansion,
we impose a selfconsistency constraint such that the operators
$\hat{\gamma}_i$, corresponding to the c-number master variables $\gamma_i$,
fulfill
\begin{equation}
  \label{selfcon} 
  \gamma_i=\langle{\hat{\gamma}_i}\rangle=
  {\rm Tr}[\hat{\gamma}_i \boldsymbol{\rho}]
  ={\rm Tr}[\hat{\gamma}_i \boldsymbol{\sigma}^{(0)}_{\{\gamma\}}].
\end{equation}
As far as the master variables are
concerned, we choose the mean-field
$\alpha^{\phantom{\dagger}}_{\mathbf{x}}$, the normal fluctuations of
the single-particle density $\tilde{f}$ and the fluctuations of the anomalous
two-particle correlation function $\tilde{m}$ such that
\begin{eqnarray}
\label{mavara}
 \langle\hat{a}^{\phantom{\dagger}}_{\mathbf{x}} \rangle=
\alpha^{\phantom{\dagger}}_{\mathbf{x}}
\quad
  f^{(c)}_{\mathbf{x},\mathbf{y}} = 
  \alpha^{\phantom{\dagger}}_{\mathbf{x}}\alpha^*_{\mathbf{y}}, 
  \quad
  m^{(c)}_{\mathbf{x},\mathbf{y}}=  
  \alpha^{\phantom{\dagger}}_{\mathbf{x}}
  \alpha^{\phantom{\dagger}}_{\mathbf{y}},\\
\label{mavarb}
  \langle\hat{a}^{\dagger}_{\mathbf{y}}
  \hat{a}^{\phantom{\dagger}}_{\mathbf{x}}\rangle=
  f^{(c)}_{\mathbf{x},\mathbf{y}} +
  \tilde{f}^{\phantom{\dagger}}_{\mathbf{x},\mathbf{y}}, 
  \quad
  \langle\hat{a}^{\phantom{\dagger}}_{\mathbf{y}}
  \hat{a}^{\phantom{\dagger}}_{\mathbf{x}}\rangle=
  m^{(c)}_{\mathbf{x},\mathbf{y}} +
  \tilde{m}^{\phantom{\dagger}}_{\mathbf{x},\mathbf{y}}.
\end{eqnarray}
A Gaussian operator $\boldsymbol{\sigma}^{(0)}_{\{\gamma\}}$ is compatible
with the requirements of (\ref{selfcon},\ref{mavara},\ref{mavarb}). In turn,
this implies the factorizability of multi operator products (Wick's theorem)
and also yields non-Gaussian corrections by calculating the contribution of
$\boldsymbol{\sigma}^{(1)}_{\{\gamma\}}$.

By studying the coordinate transformation properties of the fluctuations
\cite{Blaizot_}, one finds that the averages $\tilde{f}$ and $\tilde{m}$ are
components of a positive semi-definite, generalized density matrix $G \ge 0$ .
Thus, the system is described by a row vector $\chi$, containing the
mean-field $\alpha$ as well as its complex conjugate, and by the density
matrix $G$
\begin{equation}
\label{Gmat}
  \chi^{\phantom{\dagger}}_{\bf{x}} = 
\left(
  \begin{array}{c}
   \alpha^{\phantom{*}}_{\mathbf{x}} \\
    \alpha^{*}_{\mathbf{x}}
  \end{array}
 \right)
,\quad 
  G_{\mathbf{x,y}} = 
\left(
  \begin{array}{cc}
    \tilde{f}^{\phantom{\dagger}}_{\mathbf{x,y}} & 
    \tilde{m}^{\phantom{\dagger}}_{\mathbf{x,y}} \\
    \tilde{m}^{\phantom{\dagger}*}_{\mathbf{x,y}} & 
    \delta_{\mathbf{x},\mathbf{y}} + 
    \tilde{f}^{\phantom{\dagger}*}_{\mathbf{x,y}}
  \end{array}
 \right) .
\end{equation}
It can be shown from a Cauchy-Schwartz inequality that at $T=0$, the
generalized density matrix $G$ obeys an idem-potency relation
\begin{equation}
  \label{idempot}
  G \sigma_3 G + G = 0,\quad   
  \sigma_3=
\left(
  \begin{array}{cc}
   \mathds{1} & \phantom{-}0\\
    0& -\mathds{1}
  \end{array}
 \right).
\end{equation}
Starting with the Heisenberg equation of motion, it is straightforward to
derive the equations of motion for $\chi$ and $G$.  In order to obtain
higher-order correlation functions within the present approximation scheme
\cite{Walser99,Walser01}, like $g^{(2)}$ or $g^{(3)}$ of
(\ref{g1g2},\ref{g3}), one has to evaluate the Gaussian ${\rm Tr}[ \ldots
\boldsymbol{\sigma}^{(0)}_{\{\gamma\}}]$ as well as the non-Gaussian
contributions ${\rm Tr}[\ldots \boldsymbol{\sigma}^{(1)}_{\{\gamma\}}]$.
However it is clear that the Gaussian contribution will dominate for weak
correlations.  Thus, we have evaluated in here only the Gaussian
contributions.  But already for $\gamma \approx 1$, deviations from that can be
noticed.

\subsection{Reduction to a quasi one-dimensional, stationary configuration}
In a very prolate trap, the transverse motion in the directions $y$ and $z$ is
effectively frozen out and only amplitudes proportional to the ground state 
\begin{equation}
  \varphi_0(y,z,t) = \frac{ e^{-(y^2+z^2)/2a^2_\perp
      -i\omega_{\perp}t}}{\sqrt{\pi} a_\perp} 
\end{equation} 
need to be considered. By projecting all three-dimensional functions onto the
longitudinal axis $\mathbf{x} \rightarrow x$, one obtains the time-dependent
Hartree-Fock-Bogoliubov equations (THFB) for $\chi_x$ and $G_{x,x^\prime}$
\begin{equation}
  \label{eqm_beg}
  i\hbar \partial_t\chi_{x}= 
  \Pi_{x} \chi_{x}+\mathcal{O}(V_{bin}^2),\,\,
  i\hbar \partial_t G_{x,x^\prime}= 
  \Sigma^{\phantom{\dagger}}_{x}G^{\phantom{\dagger}}_{x,x^\prime} 
  - \mbox{h. c.}+\mathcal{O}(V_{bin}^2),
\end{equation}
with the following abbreviations for the single particle Hamiltonian and the
self energies
\begin{eqnarray}
  \label{Sigma_Mat1}
  \Pi_{x}  = 
  \left(
    \begin{array}{cc}
      \phantom{-}\Pi_N & \phantom{-}\Pi_A \\
      -\Pi_A^* & - \Pi_N^*  
    \end{array}
  \right)
,\quad 
  \Sigma_{x}  = 
  \left(
    \begin{array}{cc}
      \phantom{-}\Sigma_N & \phantom{-}\Sigma_A \\
      -\Sigma_A^* & - \Sigma_N^* \\
    \end{array}
  \right),
\end{eqnarray}
\begin{eqnarray}
  \Pi_N & =& \mathcal{H}^{\phantom{\dagger}}_{x} + 
  g f^{(c)}_{x,x} +2g  \tilde{f}^{\phantom{\dagger}}_{x,x}, 
  \quad
  \Sigma_N  = \mathcal{H}^{\phantom{\dagger}}_{x} + 
  2 g f^{(c)}_{x,x} +
  2g \tilde{f}^{\phantom{\dagger}}_{x,x},\\
  \label{hoonedim}
  \Sigma_A & =& g m^{(c)}_{x,x} +  g \tilde{m}^{\phantom{\dagger}}_{x,x},
  \quad  \Pi_A = g  \tilde{m}^{\phantom{\dagger}}_{x,x},
  \quad
  \mathcal{H}_{x} =
-\frac{\hbar^2}{2 m}\partial^2_x+\frac{1}{2\beta^2}m \omega_\perp^2 x^2.
\end{eqnarray}
In the course of the dimensional reduction, we had to introduce an effective
one-dimensional coupling constant $g = 2\hbar \omega_\perp {a}_s$, which is in
agreement with previous derivations \cite{walser04,Olshanii98, MenStrin02}.
In order to obtain the stationary solution for the time-independent 
fields $ \chi_{x}$ and $ G_{x,x^\prime}$, we make the
ansatz
\begin{equation}
  \chi(t)  =   e^{-\frac{i}{\hbar} \mu t \sigma_3}   \chi,\quad
  G(t)= e^{-\frac{i}{\hbar} \mu t\sigma_3}\, G\,
  e^{\frac{i}{\hbar} \mu t \sigma_3}
\end{equation}
which introduces the chemical potential $\mu$ and employs the Pauli matrix
$\sigma_3$ of (\ref{idempot}). This ansatz implies that the normal
fluctuations $\tilde{f}_{x,x^\prime}(t)$ become time-independent, while the
anomalous fluctuations $\tilde{m}_{x,x^\prime}(t)$ oscillate with twice the
chemical potential.  The properties of the resulting stationary HFB equation
will be investigated in section \ref{secstathfb} in the case of a homogeneous
gas of bosons and in section \ref{zerofinite} for a harmonic trapping
potential, both at zero and for finite temperatures.

The fact that the eigenvalue in the resulting stationary equations is indeed
the chemical potential \cite{Blaizot_} can be seen from a variation of the
total energy functional $ \mathcal{E}(\alpha,\tilde{f},\tilde{m}) = \langle
\hat{H} \rangle$ given by
\begin{eqnarray}
  \label{eng_fun}
  \mathcal{E}&=&\int{\rm d}x {\rm d}y \, \delta(x-y)[\alpha^{*}_y 
    (\mathcal{H}^{\phantom{\dagger}}_{{x}}+ 
      \frac{g}{2}|\alpha^{\phantom{\dagger}}_x|^2) 
    \alpha^{\phantom{\dagger}}_x +( 
      \mathcal{H}^{\phantom{\dagger}}_{{y}}+
      g\tilde{f}^{\phantom{\dagger}}_{{x},{y}})
    \tilde{f}^{\phantom{\dagger}}_{{x},{y}}] \nonumber\\
  &&+ \frac{g}{2} \int {\rm d}x [(
      2|\alpha^{\phantom{\dagger}}_x|^2 
      \tilde{f}^{\phantom{\dagger}}_{{x},{x}} 
      +  {\alpha_x}^2\tilde{m}^{*}_{{x},{x}} +  
      \frac{1}{2} |\tilde{m}^{\phantom{\dagger}}_{{x},{x}}|^2)+
    \mbox{h. c.}]+\mathcal{O}(V_{bin}^2),
\end{eqnarray}
with the constraint that the number of particles $N=\int{\rm d}x
(f^{(c)}_{x,x} + \tilde{f}_{{x},{x}})$.

\section{Analytic solution for the stationary HFB-equations in the 
homogeneous system at zero temperature}
\label{secstathfb}
For the calculations that are presented in the following sections, we use
standard parameters for $^{87}$Rb in  natural units of length
$a_\perp$ and energy $\varepsilon_\perp$
\begin{equation} 
  \begin{array}{lll}
    m  = 1.4432 \cdot 10^{-25} \mbox{ kg},& 
    a_s = 5.8209 \cdot 10^{-9} \mbox{ m}, \\
    \label{par1} 
    \omega_\perp = 2\pi \cdot800 \mbox{ Hz},& 
    \omega = 2\pi \cdot 3 \mbox{ Hz},\\
    a_\perp=  3.8128\cdot 10^{-7} \mbox{ m}, &
    \tilde{g}=2 a_s/a_\perp=3.0533 \cdot 10^{-2}. 
  \end{array}
\end{equation}

To reach the homogeneous case, we have to decrease the influence of the
external trapping potential in (\ref{hoonedim}) by weakening it to the
limit $\beta\gg 1$, where we can neglect it practically. Therefore, the
equilibrium state should possess the same translation symmetry as the generator
of the dynamics.  Consequently, we can assume that the mean field is space
independent $\alpha_x=\alpha$ and the density matrix only depends on relative
differences $r=x-x^\prime$
\begin{equation}
  \label{fourier}
  \chi_x=\chi, \qquad
  G_{x,x^\prime}= G_{x-x^\prime}=
  \int_{-\infty}^{\infty}  \frac{{\rm d}k}{2\pi} \,
  e^{-ikr} \mathcal{G}_k \;.
\end{equation}
Translation invariant systems are best described in Fourier-space, which was
introduced above. We can also choose the mean-field to be real-valued by a
suitable phase rotation. This is a consequence of the global number
conservation that is built into the dynamical HFB equations \cite{Griffin96}.
As the mean-field is real-valued $f^{(c)}=m^{(c)}=\alpha^2$ , so are the
fluctuations $\tilde{f}_0=\tilde{f}_{x,x}$ and $\tilde{m}_0=\tilde{m}_{x,x}$
and with these assumptions, the normalization constraint reads
\begin{eqnarray}
  \label{nc}
  n=\frac{N}{L}=f^{(c)} + \tilde{f}_{0},
\end{eqnarray}
where $n$ is the linear particle density on a length $L$. Furthermore the THFB
equations (\ref{eqm_beg}) simplify significantly to
\begin{equation}
  \label{hfbk}
  \mu=\tilde{g}( f^{(c)} + 2\tilde{f}_0 + \tilde{m}_0),\qquad
  0= (\Sigma_k-\mu\sigma_3) \mathcal{G}_k-\mbox{h. c.}
\end{equation}
From the equation for the chemical potential, it is clear that energy and
length scales emerge. It will be beneficial to introduce such scales for
coherence ($\kc, \xic,\omc$), the pairing correlations ($\kt, \xit,\omt$), and
their weighted sums and differences as 
\begin{eqnarray}
  \kc=\xic^{-1}=\sqrt{\omc},
  \quad \kt={\xit}^{-1}=\sqrt{\omt},
  \quad k_\pm= {\xi_\pm}^{-1}=\sqrt{\omega_\pm}\\
  \omc= 4\tilde{g} f^{(c)},\qquad \omt = -4\tilde{g} \tilde{m}_0,\qquad  
  \omega_\pm=\frac{\omc\pm\omt}{2}. 
\end{eqnarray}
In particular, in the Gross-Pitaevskii regime one can assume that
$\omc\gg\omt$.  With these definitions, one finds that the self energy is
simply a $2\times 2$ matrix in $k$-space with eigenvalues $\omega_k$
\begin{equation}
  \Sigma_k-\mu \sigma_3 =   \frac{1}{2}
\left(
  \begin{array}{cc}
    k^2 +\omp & \omm \\
    -\omm & - k^2 -\omp
  \end{array}
  \right)
,\,  \omk  =  \frac{1}{2}\sqrt{(k^2+ \omc)(k^2+\omt)}.
\end{equation}
Now, we can finally evaluate the density matrix part of the HFB equations
(\ref{hfbk}).  Moreover, we also have to consider the idem-potency relation of
(\ref{idempot}).  It holds for the vacuum state at zero temperature and one
obtains another, now quadratic relation between normal and anomalous
fluctuations in $k$-space
\begin{equation}
  \label{mFT_prem}
  \tilde{m}_k  = - \frac{\omm}{k^2+\omp}
  \left(\frac{1}{2}+\tilde{f}_k\right),\quad
  \tilde{f}_k= \tilde{m}_k^2-\tilde{f}_k^2.
\end{equation}
The system of equations can be solved point wise in $k$-space and leads to two
solutions. One of which has to be rejected on physical grounds. Thus, we find
\begin{equation}
  \label{mFT}
  \tilde{m}_k   =  -  \frac{\omm}{4 \omk},\quad
  \tilde{f}_k   =  \frac{k^2 + \omp}{4 \omk}-\frac{1}{2}. 
\end{equation}
The high momentum tail  of the correlation functions is  responsible for the
short scale behaviour in real space. In the limit $k \to \infty$ the leading
terms are 
\begin{equation}
  \tilde{m}_k \sim -\frac{\omm}{2k^2}, \quad 
  \tilde{f}_k \sim  \tilde{m}_k^2.
\end{equation}

\subsection{Diagonal contributions of normal and anomalous fluctuations}
\label{exact_mf}
In order to obtain the diagonal part of the translation invariant correlation
functions $\tilde{m}_{r}$ and $\tilde{f}_{r}$ at $r=0$,
we have to evaluate the inverse Fourier transform of (\ref{fourier})
\begin{equation}
  \tilde{m}_{0} = - \frac{\omm}{8\pi}  \int_{-\infty}^\infty 
  \frac{dk}{ \omk},\quad
  \tilde{f}_{0} =\int_{-\infty}^\infty \frac{dk}{ 2\pi} \tilde{f}_k.
\end{equation}
Serendipitously, this can be done exactly in terms of elliptic integrals
\cite{Abramowitz}
\begin{eqnarray}
  \label{m0_exact}
  \tilde{m}_0 & =& 
  - \frac{\kc}{4\pi} \left(1 + \frac{\tilde{m}_0}{f^{(c)}}\right)
  K\left(1+\frac{\tilde{m}_0}{f^{(c)}}\right),\\ 
  \label{f0_exact}
  \tilde{f}_0 & = & 
  \tilde{m}_0 - \frac{\kc}{2 \pi} 
  \left[E\left(1+\frac{\tilde{m}_0}{f^{(c)}}\right)- 
    K\left(1+\frac{\tilde{m}_0}{f^{(c)}}\right)\right], 
\end{eqnarray}
where $K$ and $E$ are the complete elliptic integral of the first kind and
second kind, respectively. Basic definitions are given in \ref{app_elliptic}.

The scaling properties of the correlation functions are most relevant for a
physical insight. Thus, we can study the Gross-Pitaevskii regime of weakly
correlated bosons, where $|{\tilde{m}_0}/{f^{(c)}}| \ll 1$ and use a series
expansion for the elliptic integral $(1-x)K(1-x) \approx \ln(4/\sqrt{x})$.
With this approximation we get
\begin{eqnarray}
  \label{m0_approx}
  \tilde{m}_0 & =& 
  - \frac{\sqrt{\tilde{g} f^{(c)}}}{4\pi}
  W\left(64\pi\sqrt{f^{(c)}/\tilde{g}}\right),\quad
  \tilde{f}_0 =
  - \frac{\sqrt{\tilde{g} f^{(c)}}}{\pi}-\tilde{m}_0.
\end{eqnarray}
In this explicit formula, we had to introduce the Lambert-$W$-function, which
is defined in \ref{lambertW} and an excellent asymptotic expansion is given in
terms of logarithms \cite{knuth96}. 

\begin{figure}
  \includegraphics[width=0.8\columnwidth]{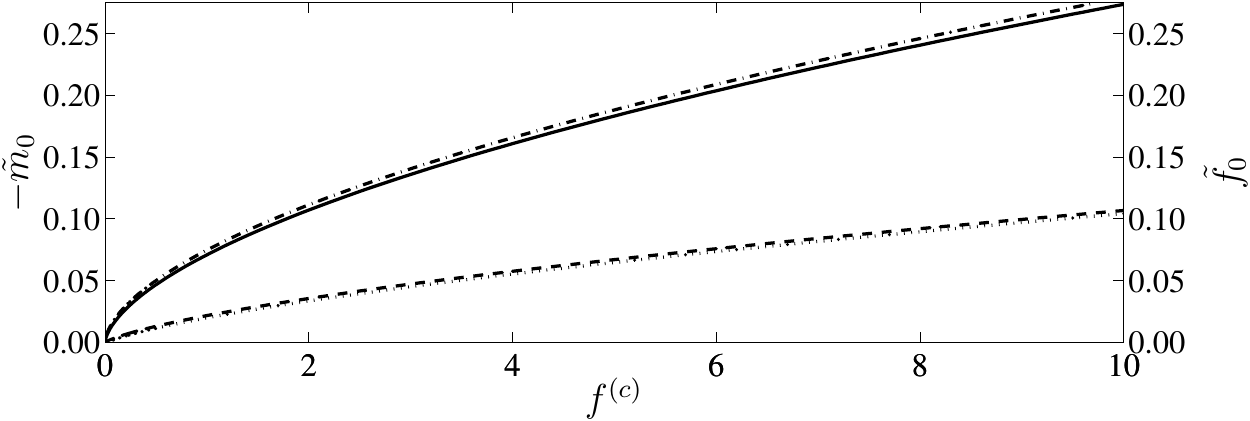}
  \caption{Diagonal part of the anomalous fluctuations $-\tilde{m}_0$ (left
    scale, solid line) and normal fluctuations $\tilde{f}_0$ (right scale,
    dashed line) versus mean field density $f^{(c)}$. The asymptotic
    approximations for $\tilde{m}_0$ (dashed dotted line) and $\tilde{f}_0$
    (dotted line) according to (\ref{m0_approx}) agree well for the
    considered parameter range.}
  \label{Comparison_m0f0}
\end{figure}

The approximations for the fluctuations
are compared to exact numerical calculations in figure \ref{Comparison_m0f0}
and give a good agreement.  We will use these approximations in the following
sections to evaluate the ground state energy and correlation functions.

\pagebreak

\subsection{Off-diagonal contribution of normal and anomalous fluctuations}

\subsubsection{Short length scale behaviour: $r \ll \xit$}
A rather simple, yet surprisingly efficient insight into the short range
behaviour of the off-diagonal of the fluctuations can be obtained by using an
iteration scheme for their Fourier transforms, which has its origin in
(\ref{mFT_prem}). Starting with $\tilde{f}_k^{(0)}=0$ and using the recursion
relation
\begin{eqnarray}
  \label{it1}
  \tilde{m}_k^{(i+1)} &=&  
  - \frac{\omm}{k^2+\omp}\left(\frac{1}{2}+\tilde{f}_k^{(i)}\right), \quad
  \tilde{f}_k^{(i+1)} =
  (\tilde{m}_k^{(i+1)})^2 - (\tilde{f}_k^{(i)})^2
\end{eqnarray}
we get a rapid  convergence towards the exact results. It is remarkable that
even with the inverse Fourier transforms of low orders of this iteration
scheme, we get a functional behaviour for $\tilde{f}(r)$ and $\tilde{m}(r)$,
which is equivalent to their exact behaviour for short ranges. However in
contrast to the exact form of the Fourier transforms of the fluctuations in
(\ref{mFT}), it is possible to perform the inverse Fourier transform
analytically in every order of the iteration scheme. A closer look reveals
that the dependence of the fluctuations on $r$ has to be of the form
\begin{eqnarray}
  \tilde{m}^{(i)}(r) &=& e^{-k_+ |r|} P^{(i)}(r),  \quad
  \tilde{f}^{(i)}(r) = e^{-k_+ |r|} Q^{(i)}(r)
\end{eqnarray}
where $P^{(i)}(r)$ and $Q^{(i)}(r)$ are polynomials in $r$ of order $2^i-2$
and $2^{i+1}-3$ respectively. Consequently the length scale on which the
correlations decay is given by
\begin{eqnarray}
  \xip &=&   \frac{1}{k_+} = 
\frac{1}{\sqrt{2 \tilde{g}(f^{(c)}-\tilde{m}_0)}}.
\end{eqnarray}
In the Gross-Pitaevskii regime, where $f^{(c)}\gg\tilde{m}_0$, we recover the
healing length $\xi \approx 1/\sqrt{2 \tilde{g}f^{(c)}}$, already introduced at
the beginning in (\ref{LLparam}).

\begin{figure}
  \includegraphics[width=0.8\columnwidth]{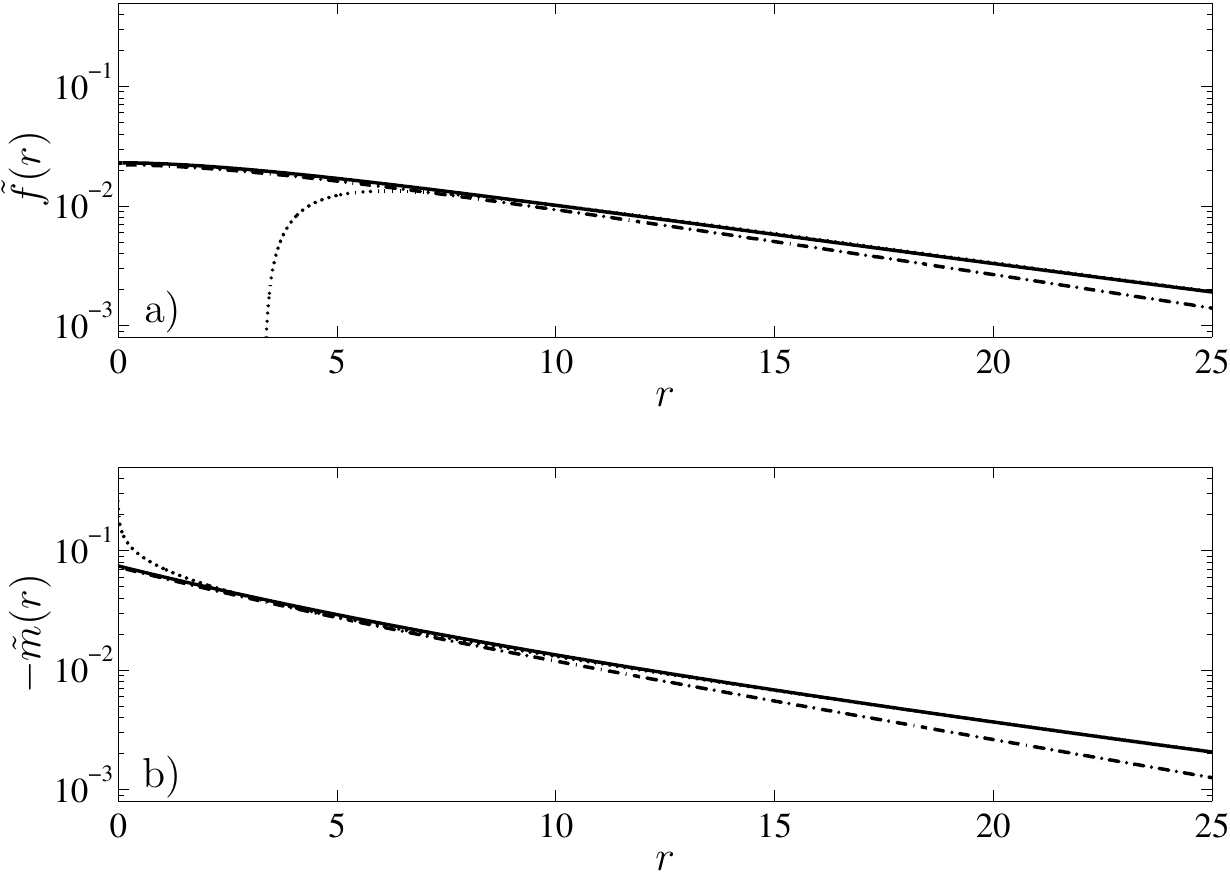}
  \caption{
    Off-diagonal part of the normal and anomalous fluctuations versus the
    distance $r$. In subplot a) we depict the normal fluctuation
    $\tilde{f}(r)$ (solid line), its short range approximation
    $\tilde{f}^{(4)}(r)$ (dashed dotted line) and the long range approximation
    (dotted line) according to (\ref{fa}). Subplot b) shows the anomalous
    fluctuation $-\tilde{m}(r)$ (solid line), the short range approximation
    $-\tilde{m}^{(4)}(r)$ thereof (dashed dotted line) and the long range
    approximation (dotted line) according to (\ref{ma}).}
  \label{mf_short_long}
\end{figure}

The short range behaviour of the anomalous fluctuation and the normal
fluctuation is depicted in figure \ref{mf_short_long}.
There, we compare the $4th$ order result of the iteration scheme to the exact
numerical evaluation of the inverse Fourier transform.  We assumed $N=100$ particles, distributed over a length of
$L=90 a_\perp$.  This length was chosen such that the density in the homogeneous
case is similar to the density in the center of the trapped system, which will
be discussed in section~\ref{zerofinite}. 
One obtains a good agreement between the approximation and the exact results
in the regime where $r \ll \xit \approx 10.5$ with  $\xi \approx 3.88$.
At the
origin we note that the anomalous fluctuation shows the typical cusp whereas
the normal fluctuation has a smooth behaviour and consequently a vanishing
first derivative at $r=0$.

\subsubsection{Long length scale behaviour: $r \gg \xit$}
\label{app_rgg}
\begin{figure}
  \includegraphics[width=0.8\columnwidth]{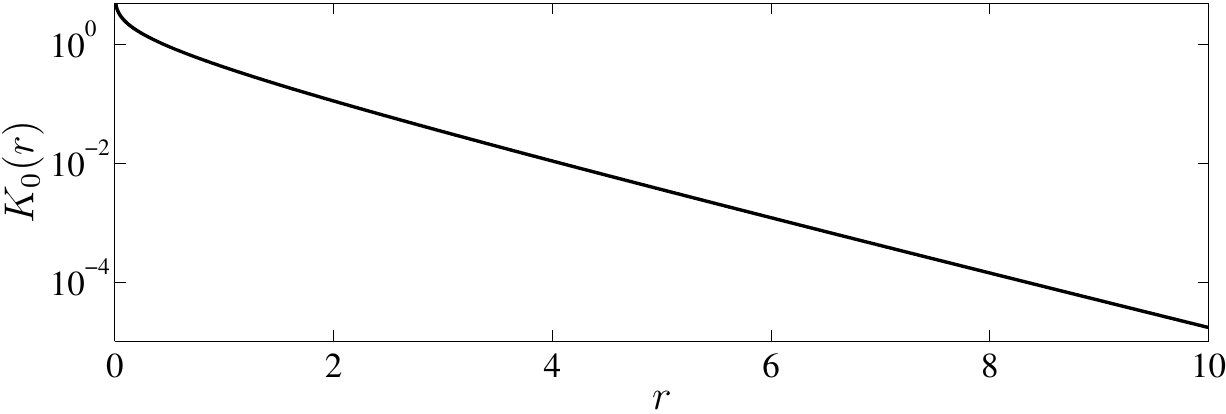}
  \caption{
    The modified Bessel function of the second kind $K_0(r)$ versus $r$. }
\label{bessel_k0}
\end{figure}

In order to get an approximation for the fluctuations in this regime, we start
with the Fourier transform of the anomalous fluctuation in (\ref{mFT}) and
note for further consideration that the Fourier transform of the modified
Bessel function of the second kind $K_0(c|r|)$ is given by
${\pi}/{\sqrt{k^2+c^2}}$.  Thus the convolution property of the Fourier
transform yields
\begin{eqnarray}
  \tilde{m}(r) &=&
  - \frac{\omm}{2\pi^2}
  \int_{-\infty}^{\infty} K_0(\kc |r'|)
  K_0(\kt |r'\!+r|)dr' .\nonumber
\end{eqnarray}
The modified Bessel function of the second kind $K_0(r)$ is presented in
figure \ref{bessel_k0}. $K_0(r)$ diverges logarithmically at the origin and
decreases exponentially for large arguments
\begin{eqnarray}
  K_0(r)\sim\left\{
    \begin{array}{ll}
      -\gamma_e + \ln \frac{2}{r} + \mathcal{O}(r^2),& r\rightarrow 0\\
      e^{-r}\left[\sqrt{\frac{\pi}{2r}} +   
        \mathcal{O}(r^{-3/2})\right],&r \to \infty
  \end{array}\right.
\end{eqnarray}
where $\gamma_e\approx0.5772$ denotes Euler's constant.

In the Gross-Pitaevskii regime where ${f^{(c)}} \gg |{\tilde{m}_0}|$, we get
from $r \gg \xit$ that also $r \gg \xic$ and therefore the first Bessel
function in the integral closely resembles a $\delta$-function.  Thus we
obtain the result
\begin{eqnarray}
  \label{ma} 
  {\tilde{m}}(r) & =& - \frac{\kc}{4\pi} 
  \left(1+\frac{\tilde{m}_0}{f^{(c)}}\right) \;K_0\left(\kt|r|\right) 
  \approx - \frac{\kc}{4\pi}  \;K_0\left(\kt|r|\right),\\
  \label{fa} 
  {\tilde{f}}(r) &= & \frac{\kc}{8\pi k_-^2}[\kc^2 K_0(\kt|r|) 
  - \kt^2 K_2(\kt|r|)] \; .
\end{eqnarray}
An alternative derivation of this result with the help of complex integration
is given in \ref{app_math}.  In figure \ref{mf_short_long} the asymptotic form
of the fluctuations in terms of the Bessel functions is compared to an exact
numerical simulation with the parameters that were mentioned in the previous
subsection. In either case the semi-logarithmic plot reveals an exponential
decay for large distances $r \gg \xit$, in agreement with the asymptotic
behaviour of the Bessel functions.

\subsection{Comparison to Lieb-Liniger theory}
\label{comp_LL}
Having all the ingredients at hand to calculate correlation functions, we are
now ready for a quantitative comparison with the results of Lieb-Liniger
theory which provides an exact solution for the behaviour of the second and
third order correlation function.

As outlined in section \ref{TDHFB_equations}, we have to obtain the values of
the correlation functions from multiple operator averages. If $\hat{o}$ is
such a general operator then
\begin{equation}
  \label{eq:oo}
  \langle\hat{o}\rangle=
  {\rm Tr}[\hat{o}\,(
  \boldsymbol{\sigma}^{(0)}_{\{\gamma\}}+
  \boldsymbol{\sigma}^{(1)}_{\{\gamma\}}+\mathcal{O}(\tilde{g}^2))]
\end{equation}
While we have already evaluated the Gaussian and non-Gaussian averages for the
multinomial operator averages \cite{Walser99}, it is clear that the Gaussian
contribution will dominate for weak correlations. 

Therefore we will focus in
here on the Gaussian contribution and disregard the non-Gaussian contributions
in the following explicit expressions  of order two and one, respectively
\begin{eqnarray}
  g^{(1)}_{x,y} & = & \frac{f^{(c)}_{{x},{y}} + 
    \tilde{f}_{{x},{y}}}{\sqrt{n_x\,n_y}}+\mathcal{O}(\tilde{g}^2),  \\
\label{g2_mv}
  g^{(2)}_{x,y} & = & 1 + 
  \frac{2\Re(f^{(c)}_{{x},{y}}
      \tilde{f}^{\phantom{*}}_{{y},{x}} + 
      {m^{(c)}_{{x},{y}}}^*\tilde{m}^{\phantom{*}}_{{y},{x}})
    +\tilde{f}^{\phantom{*}}_{{x},{y}}
    \tilde{f}^{\phantom{*}}_{{y},{x}}+ 
    \tilde{m}^{\phantom{*}}_{{x},{y}}{\tilde{m}^*_{{y},{x}}}}{n_x\,n_y}+\mathcal{O}(\tilde{g}),\\
\label{g3_mv}
  g^{(3)}_{x,y} & = & 1 + \frac{2}{n_x\,n_y}\left[2\Re
    (f^{(c)}_{{x},{y}}\tilde{f}^{\phantom{*}}_{{y},{x}} + 
      {m^{(c)}_{{x},{y}}}^*\tilde{m}^{\phantom{*}}_{{y},{x}})
    +\tilde{f}^{\phantom{*}}_{{x},{y}}
    \tilde{f}^{\phantom{*}}_{{y},{x}}+ 
    \tilde{m}^{\phantom{*}}_{{x},{y}}{\tilde{m}_{{y},{x}}}^*\right]
    \\
    &&+ \frac{1}{n_x\,n_y}\left[f^{(c)}_{{x},{x}}
    \tilde{f}^{\phantom{*}}_{{y},{y}}+\tilde{f}^{\phantom{*}}_{{x},{x}}
    \tilde{f}^{\phantom{*}}_{{y},{y}}+ 2\tilde{f}^{\phantom{*}}_{{x},{y}}
    \tilde{f}^{\phantom{*}}_{{y},{x}}+2{\tilde{m}_{{x},{y}}}^*
    \tilde{m}^{\phantom{*}}_{{y},{x}}\right] \nonumber\\
  &&+  \frac{1}{n_y^2}\;\;\left[2\Re\left(m^{(c)}_{{y},{y}}
      {\tilde{m}_{{y},{y}}}^*\right) + {\tilde{m}^{\phantom{*}}_{{y},{y}}}
    {\tilde{m}_{{y},{y}}}^* + f^{(c)}_{{y},{y}}
    \tilde{f}^{\phantom{*}}_{{y},{y}}\right] \nonumber \\
  &&+ \frac{4}{n_x\,n_y^2} \Re\left[\tilde{f}^{
      \phantom{*}}_{{x},{y}}(\tilde{m}^{\phantom{*}}_{{y},{y}}{
        \tilde{m}_{{y},{x}}}^* + {m}^{(c)}_{{y},{y}}{{\tilde{m}}_{{y},
          {x}}}^*)
    + f^{(c)}_{{x},{y}}(\tilde{m}^{
        \phantom{*}}_{{y},{y}}{\tilde{m}_{{y},{x}}}^* + 
      \tilde{f}^{\phantom{*}}_{{y},{y}}{{\tilde{f}}^{\phantom{*}}_{{y},
          {x}}})\right. \nonumber\\
  &&+ \left.m^{(c)}_{{x},{y}}(\tilde{f}^{
        \phantom{*}}_{{y},{y}}{\tilde{m}_{{y},{x}}}^* + 
      {\tilde{m}_{{y},{y}}}^*{{\tilde{f}}^{\phantom{*}}_{{y},{x}}}) 
  \right]+\mathcal{O}(\tilde{g}), \nonumber
\end{eqnarray}
where $n_x = f^{(c)}_{{x},{x}} + \tilde{f}_{{x},{x}}$ denotes the total
density. This way we can calculate the full diagonal and off-diagonal
behaviour of the correlation functions. It works equally well for the trapped
and homogeneous case.
\begin{figure}
\includegraphics[width=0.8\columnwidth]{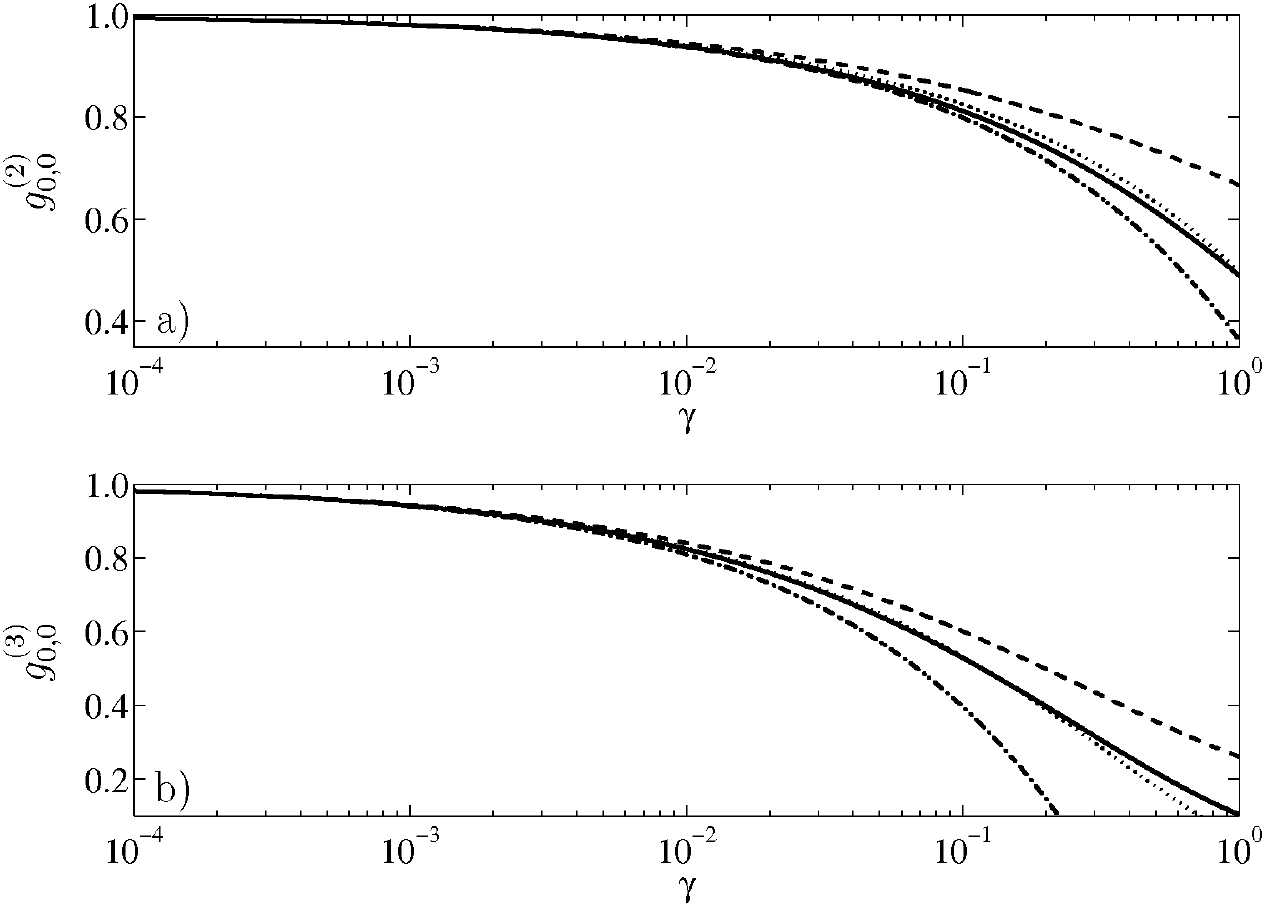}
\caption{Second and third order correlation 
  functions versus the correlation parameter $\gamma$. In subplot a) we
  compare the exact results from Lieb-Liniger theory, $g^{(2)}_{LL}$ (solid
  line), and the approximation in the GP-regime, $g^{(2)}_{LL,GP}$ (dashed
  dotted line) to $g^{(2)}_{x,x}\equiv g^{(2)}_{0,0}$ calculated with an
  extended mean-field theory. We depict exact results (dashed line) using
  (\ref{m0_exact}) and (\ref{f0_exact}) as well as approximated results
  (dotted line) using (\ref{m0_approx}). In subplot b) we depict the same
  comparison for the third order correlation function.}
\label{Comparison_g2_g3}
\end{figure}

In figure \ref{Comparison_g2_g3} we see a comparison of approximations and
exact numerical results within extended mean-field theory as well as
Lieb-Liniger theory. In either case we observe a good agreement between our
results and  Lieb-Liniger theory. However as $\gamma$
increases the deviation from the exact result grows. We attribute this 
deviation to the non-Gaussian contributions that have been dropped. 

\begin{figure}
  \includegraphics[width=0.8\columnwidth]{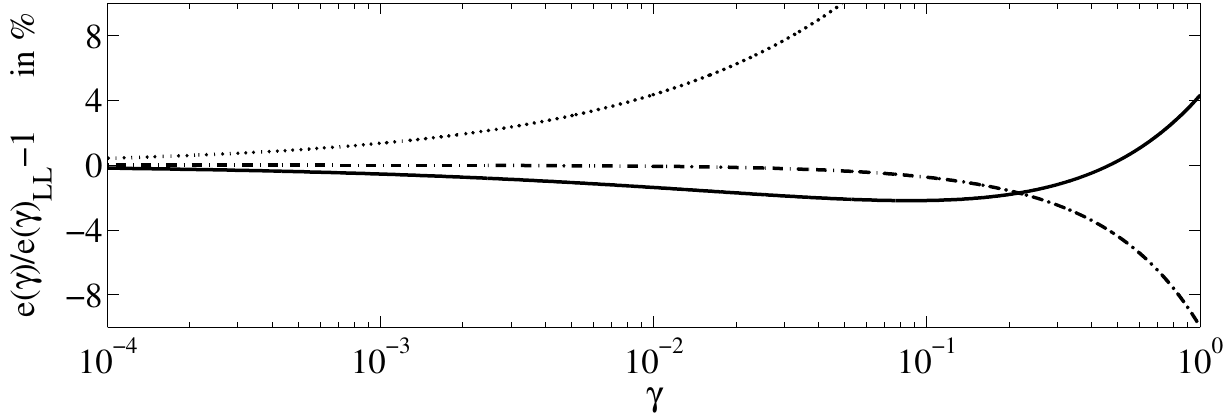}
  \caption{
    Relative deviations from the dimensionless ground state energy of
    Lieb-Liniger theory as a function of $\gamma$. Results from an extended
    mean-field approach (solid line), the simple mean-field theory without
    fluctuations (dotted line) and the Bogoliubov approach (dashed dotted
    line) are compared.}
  \label{eng_pic1}
\end{figure}

Another relevant quantity is the ground state energy of the system.  By
comparing the value of the energy functional (\ref{eng_fun}) with the
Lieb-Liniger ground state energy for a range of the correlation parameter
$\gamma$, we obtain figure \ref{eng_pic1}. In particular, we plot the relative
deviation of the ground state energies. This is to be compared with deviations
from a simple mean-field approach neglecting fluctuations and for the
Bogoliubov method in the GP-regime which includes excitations of the
mean-field. The latter approach results in \cite{Wadati02}
\begin{eqnarray}
  e(\gamma)_{{LL,GP}} & =&   
  \gamma - \frac{4}{3\pi}\gamma^{\frac{3}{2}} .
\end{eqnarray} 

In either case, we present all the results in the form of a normalized
deviation from the dimensionless ground state energy $e(\gamma)$ from
Lieb-Liniger theory given by (\ref{e_gamma}). 

\pagebreak

The results show a clear improvement over simple mean-field theory and it also
improves on the Bogoliubov method. Up to the cross-over at $\gamma \approx 1$
the maximum deviation of our results is less than 4$\%$ and we obtain reliable
results throughout the region of interest, i.e. $\gamma \leq 1$. However, this
appears to be the limit for a quasi one-dimensional extended mean-field theory
and different approaches have to be used in the strongly correlated regime.

\pagebreak

\section{Numerical results for trapped atoms at zero and finite temperature}
\label{zerofinite}
\subsection{The zero temperature limit for a trapped gas}
\label{zero}
In the previous section we have studied the homogeneous case. In here, this
will be extended to harmonically trapped systems and we present correlation
functions up to third order.  First of all we depict the spatial shape of the
master variables $\tilde{f}$, $\tilde{m}$ and of the quantities $f^{(c)}$,
$m^{(c)}$, which are essential for the calculation of the correlation
functions. The plots show numerical simulations for a particle number of
$N=10^2$ in a trap with standard parameters for $^{87}$Rb according to
(\ref{par1}).
\begin{figure}
\includegraphics[angle=-90,width=0.8\columnwidth]{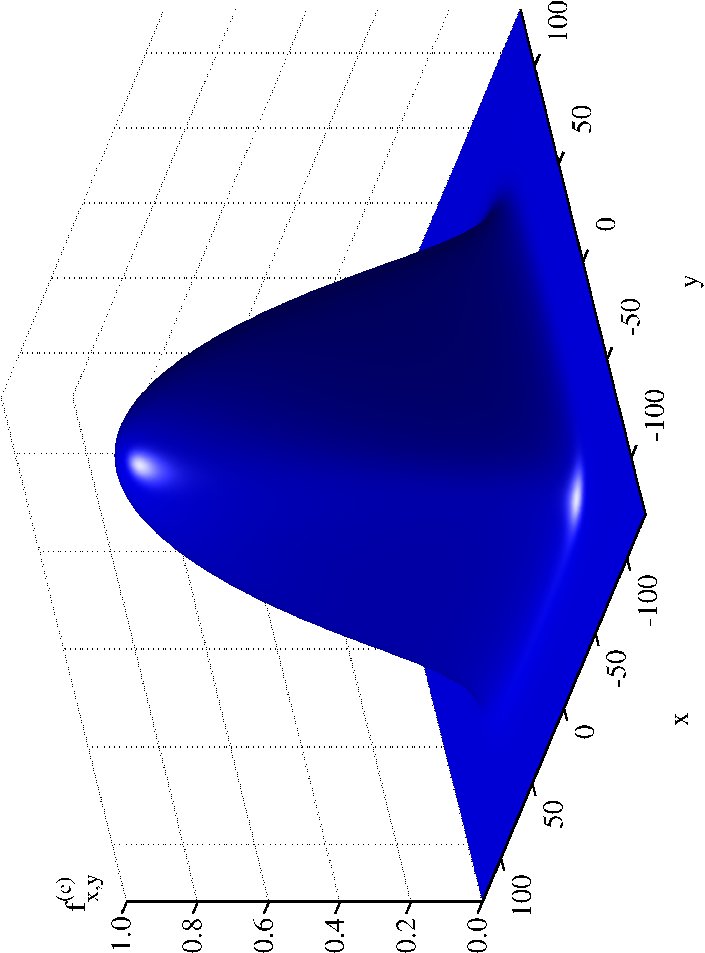}
\caption{Coherent  single particle density matrix $f^{(c)}_{x,y}$ versus
  $x$ and $y$. As the ground state is real valued, the coherent part of the
  pairing field $m^{(c)}_{x,y}$ is also represented in this figure.}
\label{fc_pic}
\end{figure}

The coherent contribution to the single particle density matrix
$f^{(c)}_{x,y}$ in figure \ref{fc_pic} has off-diagonal long range order and 
extends over the complete system. As the Hamiltonian for a one-dimensional
trap is real-valued, so is the ground state solution $\alpha_x$. Hence, the
coherent contribution of the pairing field $m^{(c)}_{x,y}$ is identical to
$f^{(c)}_{x,y}$ and shown in figure \ref{fc_pic}.

\begin{figure}
\includegraphics[angle=-90,width=0.8\columnwidth]{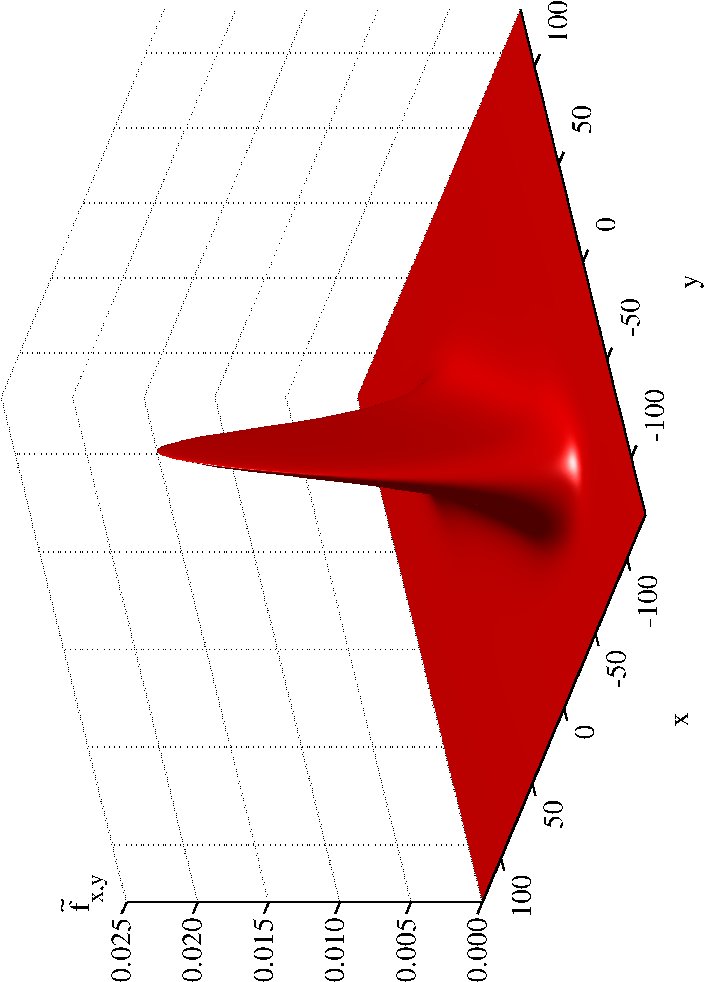}
\caption{Normal fluctuations $\tilde{f}_{x,y}$ versus $x$ and $y$.} 
\label{fsq_pic}
\end{figure}

\begin{figure}
\includegraphics[angle=-90,width=0.8\columnwidth]{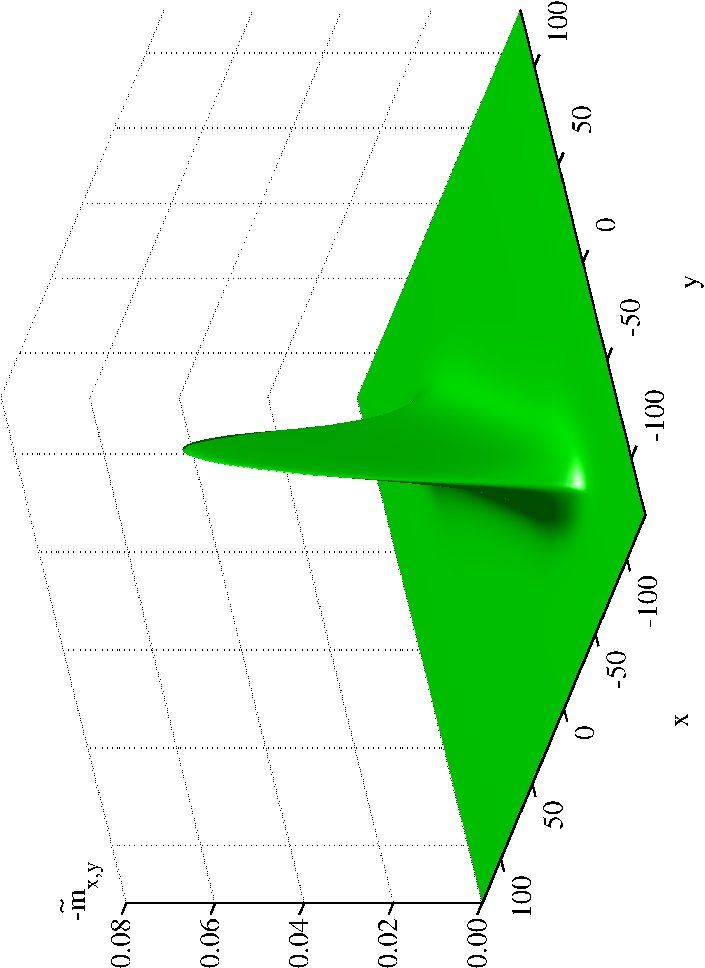}
\caption{Anomalous fluctuations $-\tilde{m}_{x,y}$ versus $x$ and $y$.} 
\label{msq_pic}
\end{figure}

In contrast to the coherent contributions, the normal fluctuation
$\tilde{f}_{x,y}$ in figure \ref{fsq_pic} and the anomalous fluctuation
$\tilde{m}_{x,y}$ in figure \ref{msq_pic} are primarily localized along the
diagonal.

\pagebreak

The coherence in the off-diagonal direction is only of short range
and the negativity of the pairing field is an indication of a reduced
likelihood of finding two particles at the same location.

\begin{figure}
  \includegraphics[angle=-90,width=0.8\columnwidth]{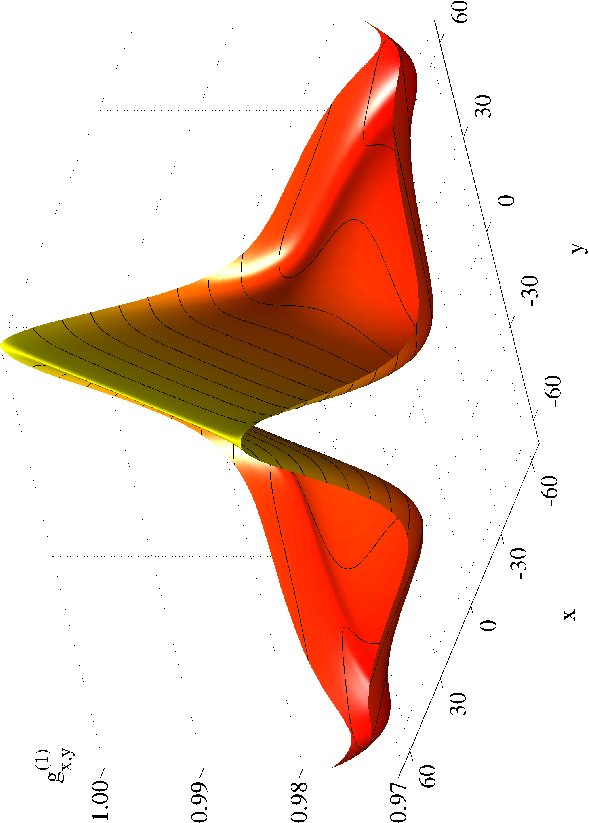}
  \caption{First order correlation 
    function $g^{(1)}_{x,y}$ versus $x$ and $y$.}
  \label{g1_trap_pic_1}
\end{figure}

\begin{figure}
  \includegraphics[angle=-90,width=0.8\columnwidth]{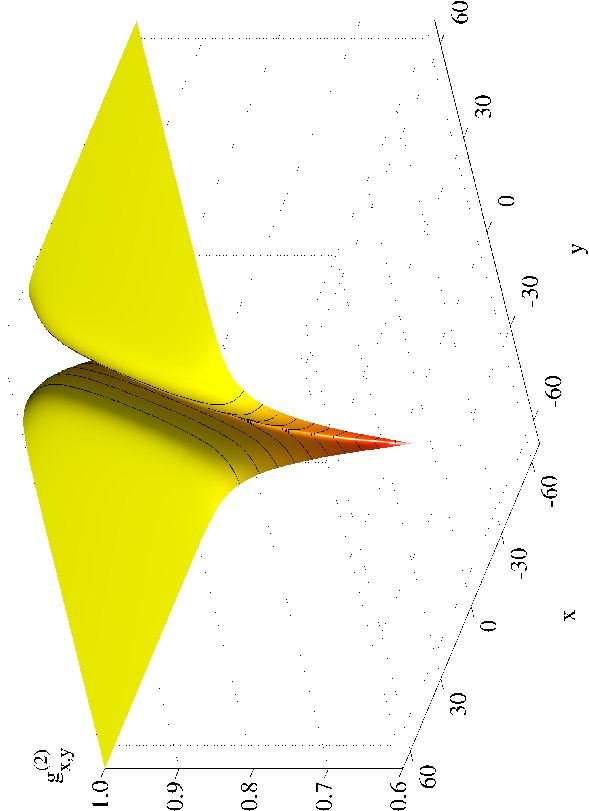}
  \caption{Second order correlation function 
    $g^{(2)}_{x,y}$ versus $x$ and $y$.} 
  \label{g2_trap_pic_1}
\end{figure} 

\begin{figure}
\includegraphics[angle=-90,width=0.8\columnwidth]{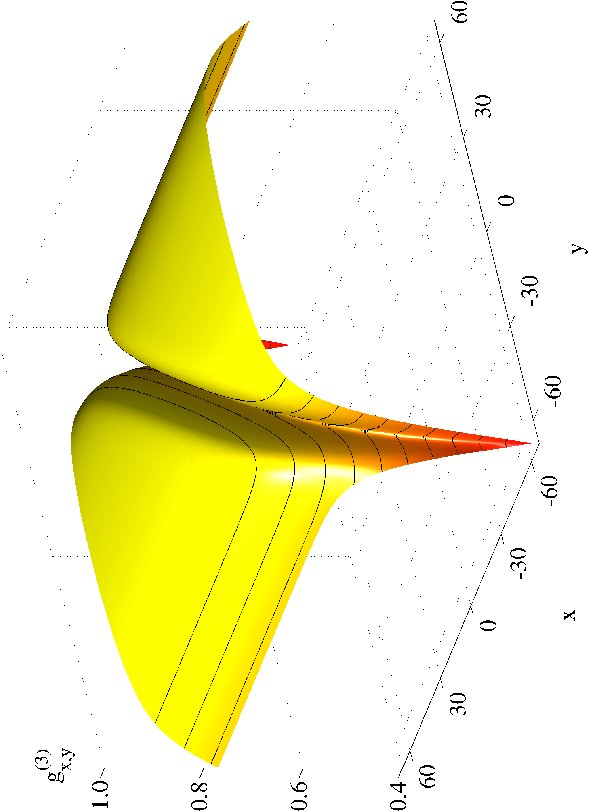}
\caption{Third order correlation function $g^{(3)}_{x,y}$ versus $x$ and $y$.} 
\label{g3_trap_pic_1}
\end{figure} 

The behaviour of the first, second and third order correlation function is
presented in figures \ref{g1_trap_pic_1}, \ref{g2_trap_pic_1} and
\ref{g3_trap_pic_1}. A generic feature of all three correlation functions is
that they become more pronounced for smaller particle numbers. For the first
order correlation function the diagonal has to be identical to one and the
deviation in the off-diagonal is fairly small as expected for a coherent
system. However, the second order density-density correlation is a more sensitive
probe as this correlation function is less than one, thus exhibits
non-classical behaviour.  This anti-bunching is particularly strong for smaller
particle numbers when we approach the Tonks-Girardeau regime of a fermionized
Bose gas and the correlation function vanishes eventually. Recently this
effect has been investigated in a number of experiments, e.g.
\cite{Weiss05, Raizen2005, Paredes2004} and confirms the
theoretical predictions.  The same statements apply to the third order
correlation function and it can be observed that the deviation from one is
even more pronounced.  This also implies that the third order correlation
function \cite{Tolra04, cornell597} is the most sensitive probe for quantum aspects of
the field. In addition we notice values which are clearly below one for $|y|\gg 1$ and $x \neq y$, because in this case $g^{(3)}_{x,y} \approx g^{(2)}_{y,y}$. This can easily be seen by looking at (\ref{g2_mv},\ref{g3_mv}) and taking into account that all terms with off-diagonal contributions of the fluctuations in $g^{(3)}_{x,y}$ are negligible for $x \neq y$.

\subsection{Behaviour in the center of the trap}

\begin{figure}
  \includegraphics[width=0.8\columnwidth]{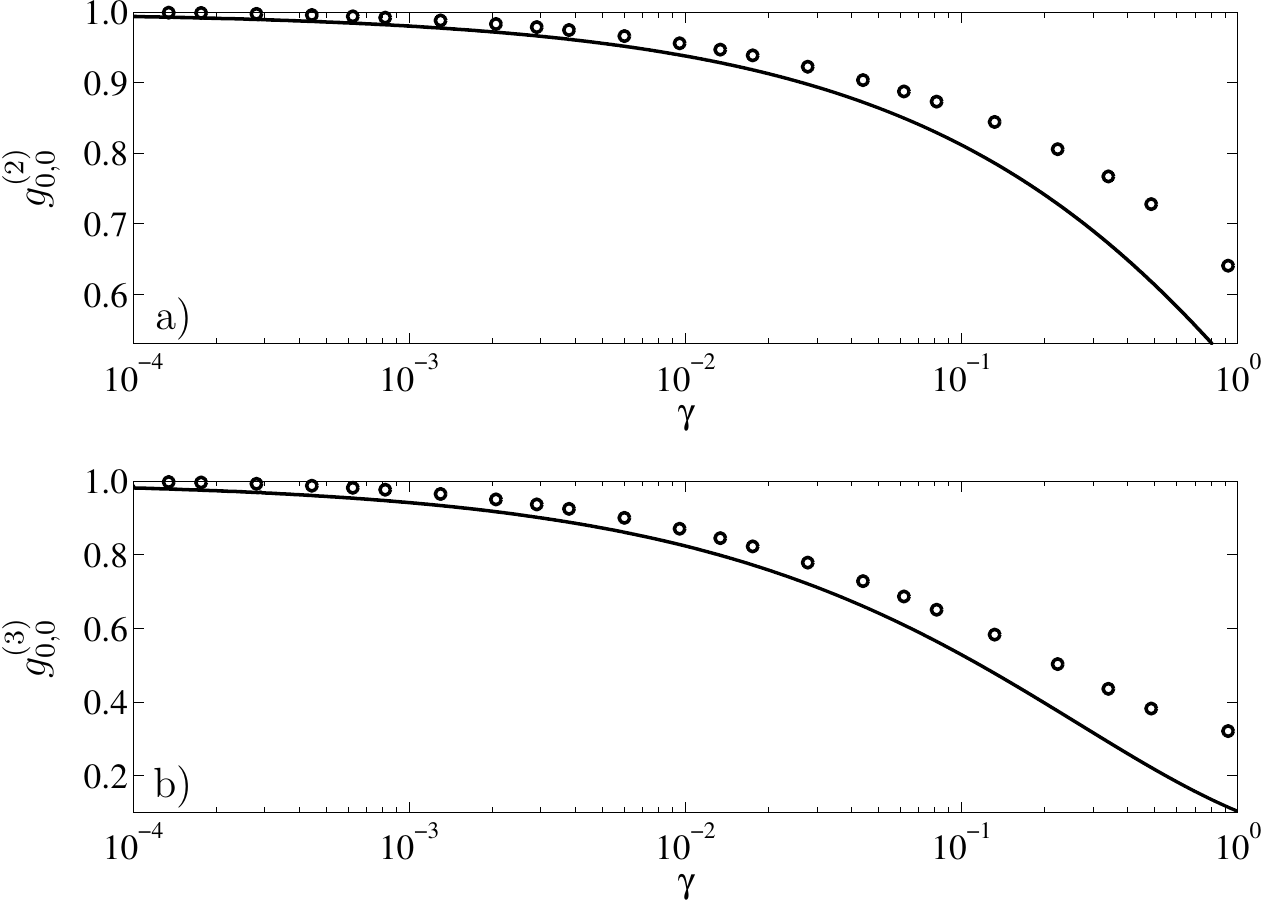}
  \caption{Correlation 
    functions versus the correlation parameter $\gamma$. In subplot a) we
    depict $g^{(2)}_{0,0}$ and compare simulation results for the trapped case
    (circles) with analytic calculations obtained with Lieb-Liniger theory
    (solid line). In subplot b) we depict $g^{(3)}_{0,0}$ and again compare
    simulation results for the trapped case (circles) with analytic
    calculations obtained with Lieb-Liniger theory (solid line). In both
    comparisons the circles originate for particle numbers $N$ ranging from
    $N=10^5$ on the left hand side to $N=10^0$ on the right hand side.}
  \label{g2_trap_orig_comp_pic}
\end{figure}

In figure \ref{g2_trap_orig_comp_pic} we
compare the results of our simulations for the second and third order
correlation function with Lieb-Liniger theory. In contrast to the comparison in subsection \ref{comp_LL} an external
potential is now included in the calculations with the extended mean-field
theory whereas the theoretical curve is for a homogeneous gas of bosons. Our
simulations are for particle numbers ranging from $N=10^0-10^5$ and we only
used the values of the correlation functions in the center of the trap for the
comparison. Compared to subsection \ref{comp_LL} the results in the
trapped case deviate slightly more from the exact results originating from the
homogeneous Lieb-Liniger theory but the qualitative behaviour is very
similar.

\subsection{Diagonal behaviour in the local density approximation}

The local density approximation (LDA) is a frequently employed approximation
scheme to transfer results of homogeneous systems to spatially trapped gases.
It is assumed that a smooth variation of the density profile can be
incorporated by an adiabatic adjustment of a locally uniform gas.  The LDA uses
a local effective chemical potential \cite{Shlyap05}
\begin{eqnarray}
  \mu(x) = \mu_0 - V(x) = \mu_0 -\frac{1}{2}m \omega^2 x^2 \; ,
\end{eqnarray}
where $\mu_0$ denotes the global equilibrium chemical potential. In order for
the LDA to be applicable, it is thus necessary that the short-range
correlation length is much smaller than the characteristic inhomogeneity
length.

\begin{figure}
  \includegraphics[width=0.8\columnwidth]{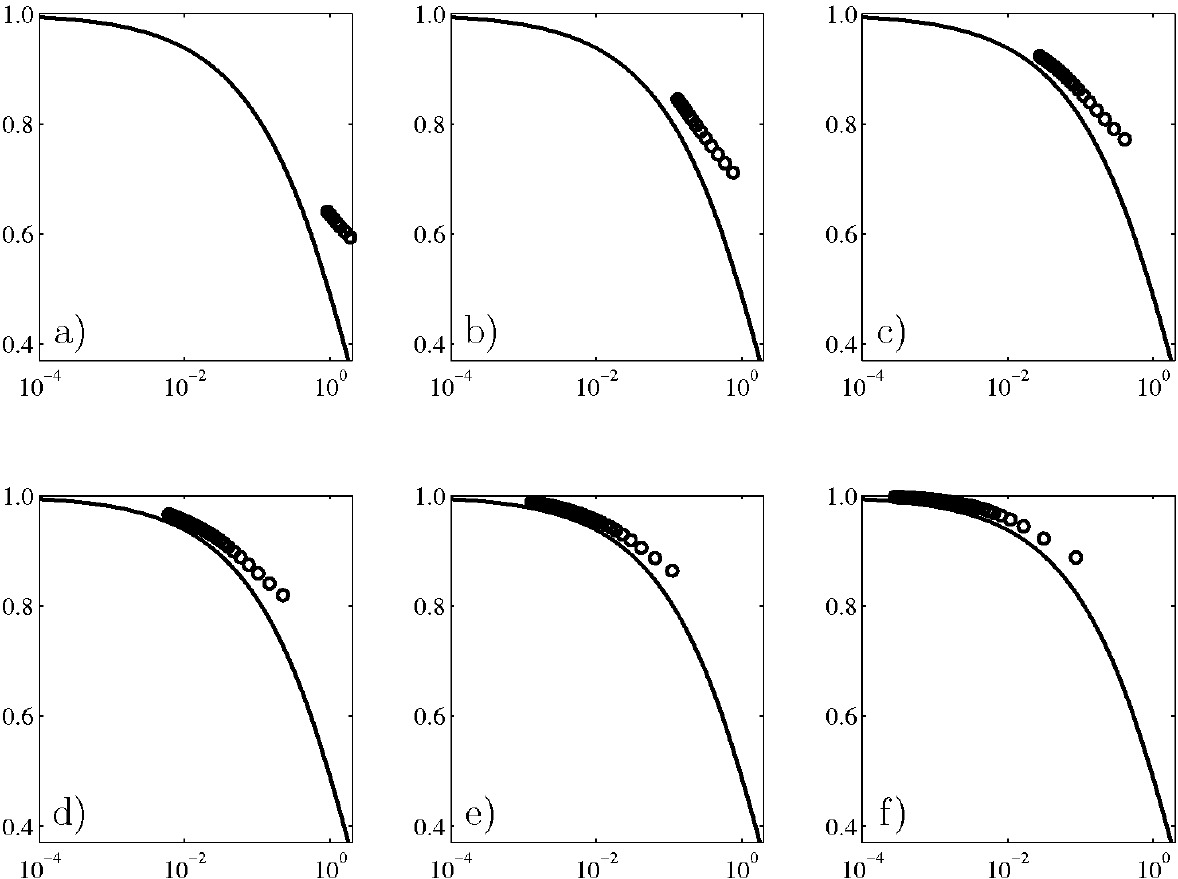}
  \caption{The diagonal
    second order correlation function $g^{(2)}_{x,x}$ versus $\gamma$ for
    various particle numbers. With an increasing value of the correlation
    parameter $\gamma$ the circles correspond to points further outwards from
    the origin. We compare results for the trapped case (circles)
    with analytic calculations for the second order correlation function
    obtained with LDA-Lieb-Liniger theory (solid line). Plots a) to c) are for
    $N=10^0$ , $N=10^1$ and $N=10^2$ (from left to right) and plots d) to f)
    are for $N=10^3$, $N=10^4$ and $N=10^5$ (from left to right).}
  \label{g2_trap_diag_comp_pic}
\end{figure}

\pagebreak

In this context, we want to compare the diagonal behaviour of our numerically
calculated correlation functions to theoretical predictions.  By definition,
the first order correlation function is identical to one along the diagonal
and our data behaves accordingly. For the second and third order
correlation function we will compare our results with the predictions from
Lieb-Liniger theory in the LDA. Naturally the LDA works best in the center of the trap. It can not be expected
to work in regions where the density drops rapidly and the inhomogeneity
length is very small in these regions. 

\begin{figure}
  \includegraphics[width=0.8\columnwidth]{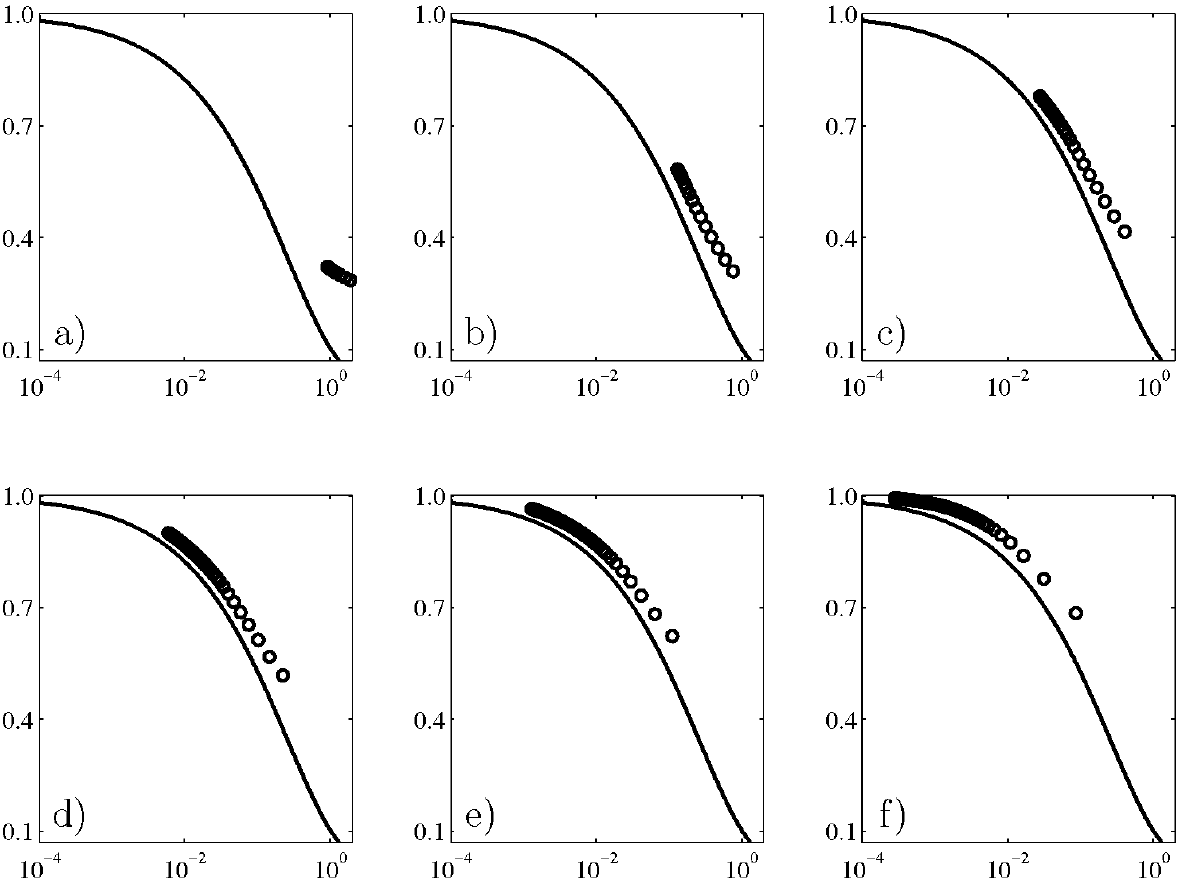}
  \caption{The diagonal 
    third order correlation function $g^{(3)}_{x,x}$ versus $\gamma$ for
    various particle numbers. With an increasing value of the correlation
    parameter $\gamma$ the circles correspond to points further outwards from
    the origin. We compare results for the trapped case (circles) with
    analytic calculations for the third order correlation function obtained
    with LDA-Lieb-Liniger theory (solid line). Plots a) to c) are for $N=10^0$
    , $N=10^1$ and $N=10^2$ (from left to right) and plots d) to f) are for
    $N=10^3$, $N=10^4$ and $N=10^5$ (from left to right).}
  \label{g3_trap_diag_comp_pic}
\end{figure}

In the Gross-Pitaevskii regime the
chemical potential $\mu$ connects the density $n$ to the correlation parameter
$\gamma$, via
\begin{eqnarray}
  \label{GP_mu}
  \mu(x) = g n(x),\quad \gamma(x)= \frac{m g}{\hbar^2 n(x)}.
\end{eqnarray}

In our simulations we tune the particle number in the trap, which decreases
$\gamma$ for an increasing number of particles.  Qualitatively one can expect
that the inhomogeneous correlation functions are higher than the homogeneous
results because in the LDA the external potential leads to a smaller chemical
potential and according to (\ref{GP_mu}) also to a smaller density compared to
the homogeneous case.  Due to the monotonous decrease of the correlation
functions, there is a tendency of the inhomogeneous values to be shifted to
larger $\gamma$ values.  All the features that have just been described can be
seen in figures \ref{g2_trap_diag_comp_pic} and \ref{g3_trap_diag_comp_pic}, where
we plotted the correlation functions for particle numbers ranging from
$N=10^0-10^5$ and restricted the plotted regions to the Thomas-Fermi radius.

\pagebreak

\subsection{The finite temperature result for a trapped gas}
\label{finite}
The zero-temperature results of the previous section can be extended easily to
account for finite temperature effects \cite{walser04,Blaizot_}.  One obtains
an equilibrium solution for the density matrix $G$ of the thermal system
(\ref{Gmat}) from the eigenstates of the selfenergy matrix
(\ref{Sigma_Mat1}), according to the Bose-Einstein distribution. We present
results in the present section for a particle number of $N=100$ and a
temperature $T=10\;\hbar \omega/k_B$.

\begin{figure}
  \includegraphics[angle=-90,
  width=0.8\columnwidth]{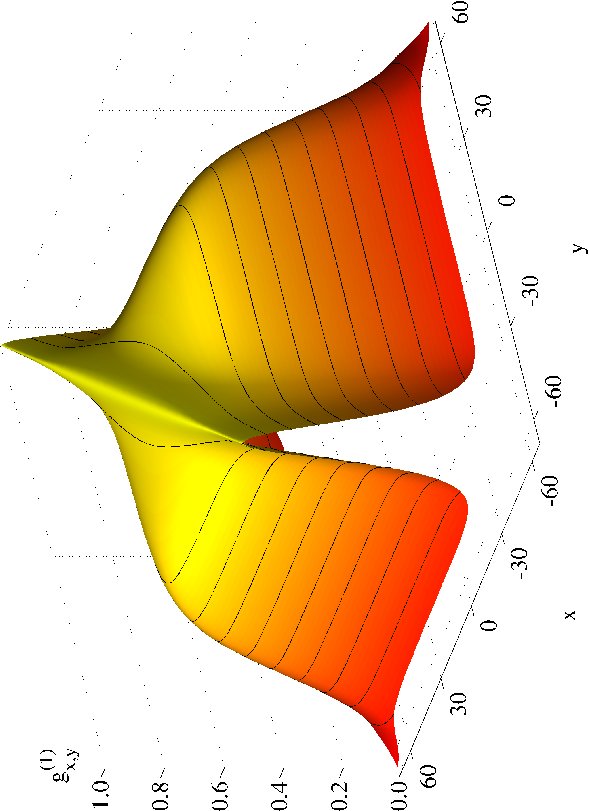}
  \caption{Finite temperature first order correlation function 
$g^{(1)}_{x,y}$ versus $x$ and 
    $y$, for $N=100$ and $T=10\;\hbar \omega/k_B$.  }
  \label{g1_trap_pic_T}
\end{figure}

\begin{figure}
  \includegraphics[angle=-90,
  width=0.8\columnwidth]{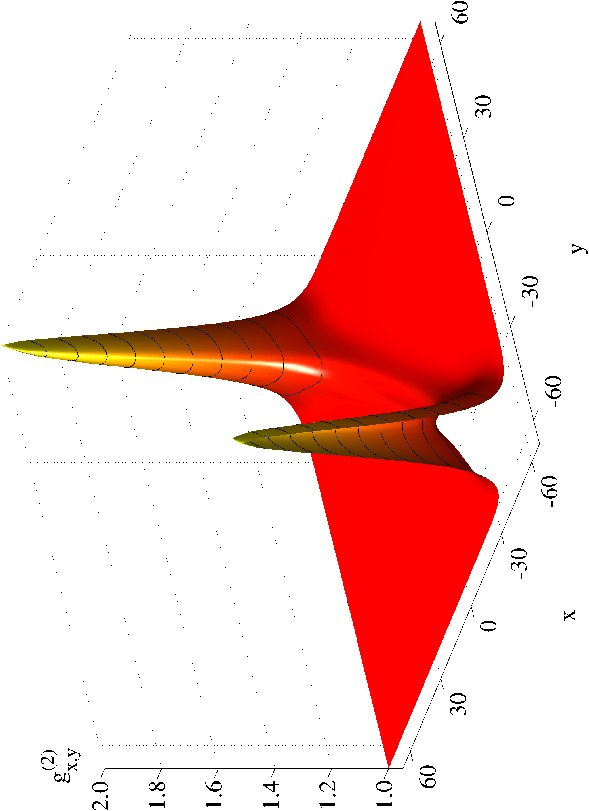}
  \caption{Finite temperature second order correlation function 
    $g^{(2)}_{x,y}$ versus $x$ and $y$, for $N=100$ and $T=10\;\hbar
    \omega/k_B$.                              } 
  \label{g2_trap_pic_T}
\end{figure}

\begin{figure}
  \includegraphics[angle=-90,
  width=0.8\columnwidth]{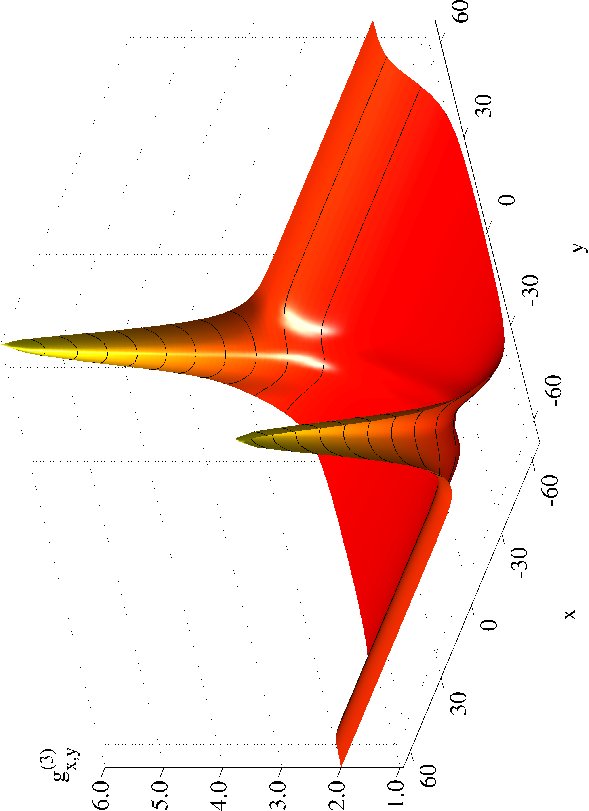}
  \caption{Finite temperature third order correlation function $g^{(3)}_{x,y}$ 
    versus $x$ and $y$,  for $N=100$ and $T=10\;\hbar
    \omega/k_B$.} 
  \label{g3_trap_pic_T}
\end{figure} 

The main thermal effect is a strong increase of the fluctuations at the edge
of the trap at the cost of a reduction of the condensate density
\cite{Hutchinson97}. This effect is clearly seen by comparing the first order
correlation function in figure \ref{g1_trap_pic_T} to the zero temperature
result in figure \ref{g1_trap_pic_1}. At finite temperatures, we also obtain a
reduction of first order coherence. Consequently, this leads to a situation
where the gas is almost thermalized at the edge of the trap, whereas it is
coherent in the center. The suppression of density fluctuations, also known as
anti-bunching, is also less pronounced at finite temperature.  This can be seen by
comparing figures \ref{g2_trap_pic_T} and \ref{g3_trap_pic_T} to figures
\ref{g2_trap_pic_1} and \ref{g3_trap_pic_1}, which give the zero temperature
results.

\pagebreak

For a thermal gas of noninteracting bosons, one finds $g^{(2)}_{0,0}=2!$ and
$g^{(3)}_{0,0}=3!$. It can be seen that these values are attained at the edge
of the trap where fluctuations dominate. In figure \ref{g3_trap_pic_T} we also
notice a value of two for $|y|\gg 1$ and $x \neq y$, because we again have
$g^{(3)}_{x,y} \approx g^{(2)}_{y,y}=2$ in this case.

\begin{figure}
  \includegraphics[width=0.8\columnwidth]{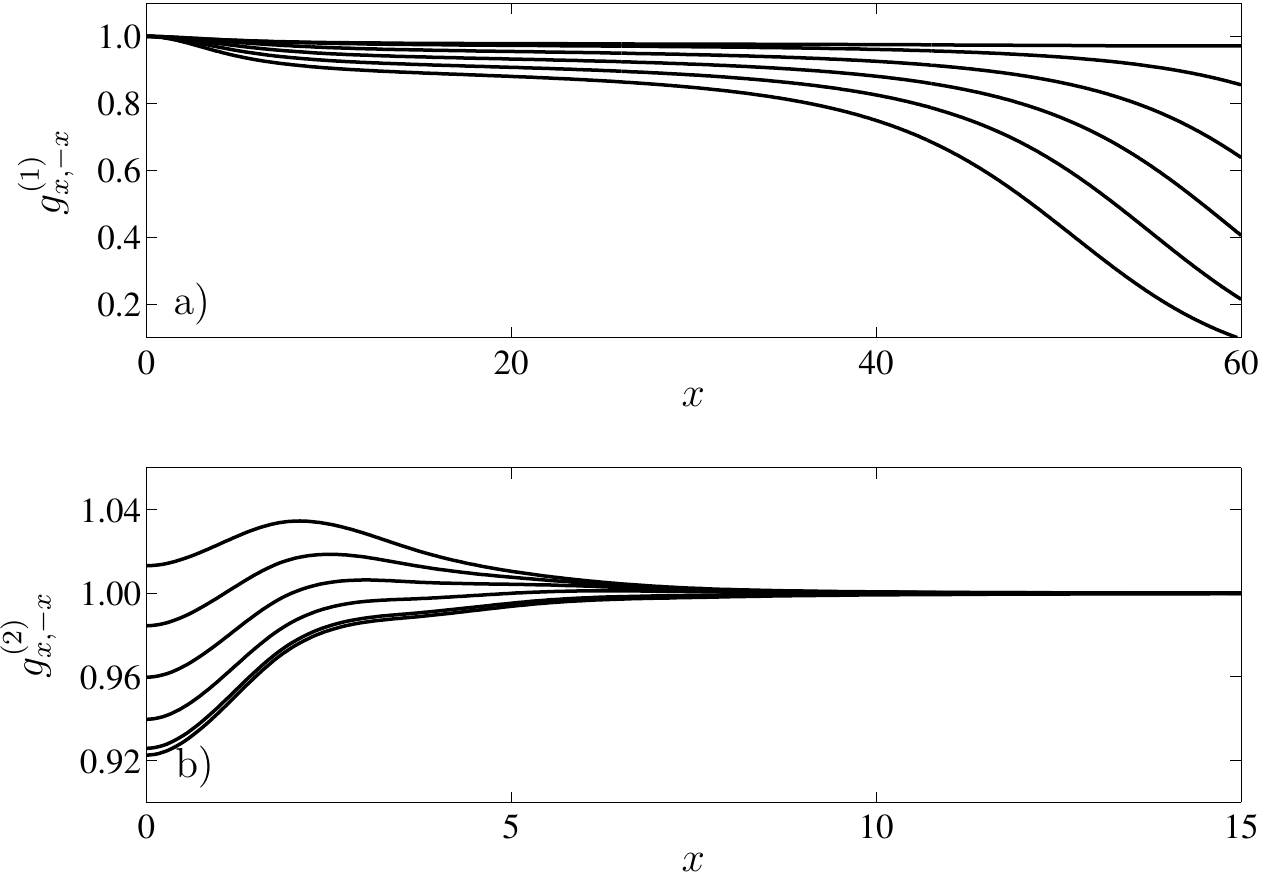}
  \caption{Off-diagonal first $g^{(1)}_{x,-x}$ (a) and second 
    order correlation function $g^{(2)}_{x,-x}$ (b) versus $x$.  The
    individual curves correspond to temperatures $k_B T = 0\,
    \hbar \omega$ (smallest value for $x=0$) to $10\, \hbar \omega$ (largest
    value for $x=0$) with increments of $2\, \hbar \omega$.}
  \label{g1_trap_varT}
\end{figure} 

In figure \ref{g1_trap_varT}, we present the off-diagonal of the first and
second-order correlation function $g^{(1,2)}_{x,-x}$ versus $x$ for
temperatures from $k_B T = 0-10\, \hbar \omega$ with increments of $2\, \hbar
\omega$. It can be noticed that correlations are strongly attenuated with
increasing temperature.  Looking at the off-diagonal of $g^{(2)}_{x,-x}$ in
figure \ref{g1_trap_varT}, we see a reduction of the anti-bunching dip in the
center with increasing temperatures, however it is still present for high
values of the temperature.  It can be understood qualitatively from the
stronger increase of fluctuations with the temperature at the edge of the
trap. Thus, the anti-bunching dip in the center of the trap remains visible
even at finite temperatures.

\section{Conclusions and outlook}
\label{conclusion}

We have presented a detailed study of an extended mean-field theory which
shows that it is a feasible approach for the description of a weakly
correlated gas of bosons in a quasi-1D setup. This approach agrees well with
exact predictions of zero-temperature Lieb-Liniger theory and can easily be
applied to spatially inhomogeneous systems at finite temperature. We
have not analyzed the finite temperature theory of Yang-Yang due its
complexity.

There are many relevant applications for using this extended mean-field theory
in different geometrical configurations like a double-well potential
\cite{Oberthaler05,Oberthaler06a}.  Yet another extension of our approach is
the dimensional crossover out-of-equilibrium where in general the increase of
available phase space volume leads to a decrease of correlations. An
evaluation of the such correlation functions is work in progress.

\section*{Acknowledgments}
The authors acknowledge the support by the German Research Foundation via SFB/TRR 21 which is a collaboration of the Universities of Stuttgart,
T\"ubingen, Ulm and the Max Planck Institute for Solid State
Research in Stuttgart.

\pagebreak

\appendix

\section{Higher transcendental functions}
\subsection{Complete elliptic integrals}
\label{app_elliptic}

Following the definitions and the notation in \cite{Abramowitz}, the complete
elliptic integral of the first kind reads
\begin{eqnarray}
  K(m) & = & \int_0^{\pi/2} \frac{ {\rm d}\theta}{\sqrt{1-m\sin^2 \theta}},
\end{eqnarray}
where the parameter $0\leq m\leq 1$. For the calculation of
$\tilde{m}_0$ in section \ref{exact_mf}, we encounter an integral of the form
\begin{eqnarray}
  I_1 = \int_0^{\infty} \frac{{\rm d}k }{\sqrt{k^2+\omc}\sqrt{k^2+\omt}} = \frac{K(m)}{\kc}
\end{eqnarray}
where $\omc>\omt$. We can show that the evaluation of this integral leads to
the complete elliptic integral of the first kind by making the
substitution $k = \kc \cot \theta$ and using $m=(\omc-\omt)/\omc$.

Similarly the complete elliptic integral of the second kind is defined as
\begin{eqnarray}
  E(m) & = & \int_0^{\pi/2} {\sqrt{1-m\sin^2 \theta}}\, {\rm d} \theta \; .
\end{eqnarray}
In order to calculate $\tilde{f}_0$ in section \ref{exact_mf} we end up with
the integral
\begin{eqnarray}
  I_2 = \int_0^{\infty} {\rm d} k \, \frac{\sqrt{k^2+\omc}-
    \sqrt{k^2+\omt}}{\sqrt{k^2+\omt}}
\end{eqnarray}
after separating the constant contribution which leads to the previously
discussed integral. The same substitution as above $k = \kc \cot \theta$
simplifies the integral to the form
\begin{eqnarray}
  I_2 = \kc \int_0^{\pi/2} \frac{1-
    \sqrt{1-m \sin^2 \theta}}{\sin^2 \theta \sqrt{1-m \sin^2 \theta}}\, 
  {\rm d} \theta=
 - \kc \left(E(m)-K(m) \right) \; .
\end{eqnarray}

\subsection{The Lambert-W function}
\label{lambertW}
The Lambert-$W$-function is implicitly defined by the solution of the
transcendental equation \cite{knuth96}
\begin{equation}
  z = W e^W.  
\end{equation}
In the case of large arguments $z\gg 1$,
one can use an asymptotic expansion
\begin{equation}
  W(z)=\ell_1-\ell_2+\ell_2/\ell_1+\ldots,
\end{equation}
with $\ell_1=\ln z$ and $\ell_2=\ln\ln z$.  
\section{Deformation of the integration contour in the complex plane }
\label{app_math}
\begin{figure}
  \centering \leavevmode
  \includegraphics[width=0.8\columnwidth]{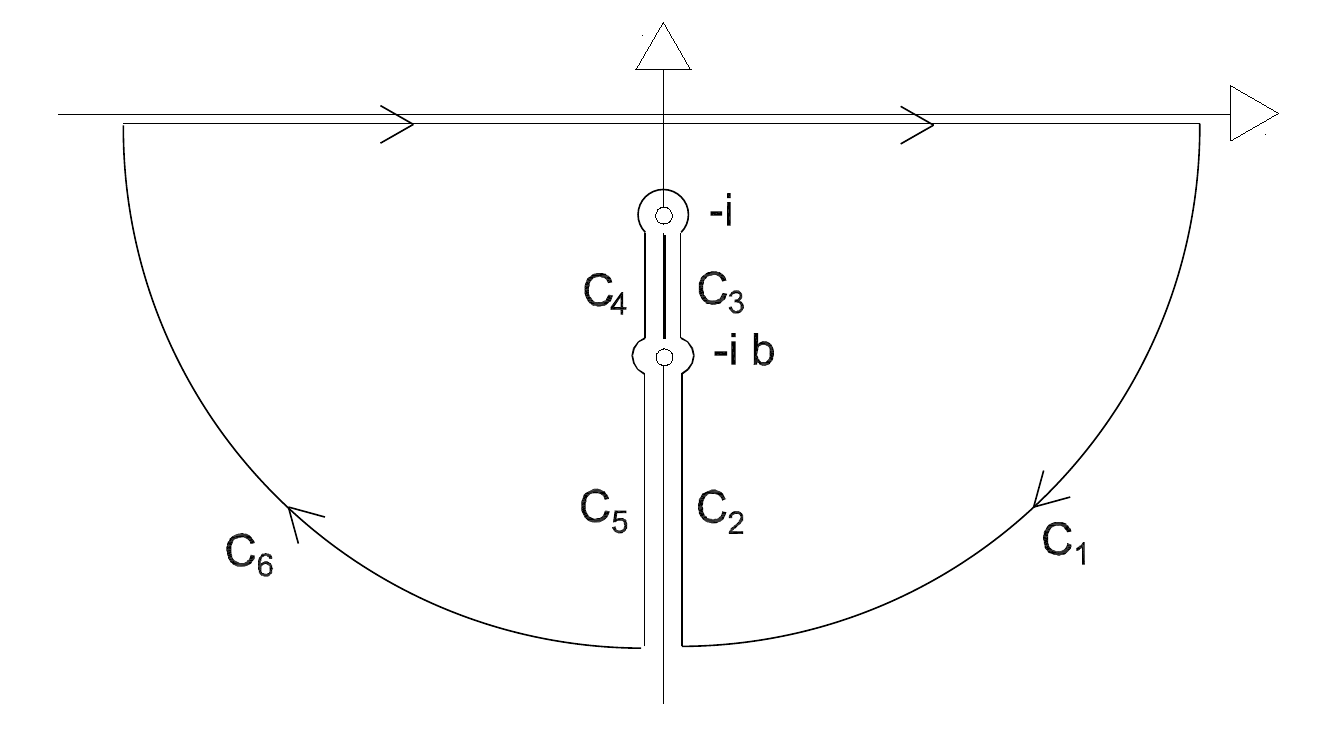}
  \caption{Integration contour for the evaluation of (\ref{exact_m}).}
  \label{contour}
\end{figure}

The results of subsection \ref{app_rgg} can be derived alternatively with the
help of complex integration \cite{Migdal}. If we take the inverse Fourier
transform of the anomalous fluctuation of (\ref{mFT}), we get
\begin{eqnarray}
  \label{exact_m}
  {\tilde{m}}\left({r}\right) =
  - \frac{ \omm }{4\pi} 
  \int_{-\infty}^\infty \frac{e^{-ikr}}{\sqrt{k^2+
      \omc}\sqrt{k^2+\omt}}\, dk \;.
\end{eqnarray}
Using the substitutions $k=\kt z$, 
$r'=\kt r$ and $b^2=\kc^2/\kt^2$ this equation reduces to
\begin{eqnarray}
  {\tilde{m}}\left({r}\right) =
  - \frac{ \omm }{4\pi} 
  \int_{-\infty}^\infty 
  \frac{e^{-ir'z}}{\sqrt{z^2+b^2}\sqrt{z^2+1}}\, \frac{dz}{\kt} \;.
\end{eqnarray}
For the evaluation of this integral we make a branch cut between $-i$ and
$-ib$ and choose the path of integration as can be seen in Fig. \ref{contour}.

The contributions from $C_1$ and $C_6$ vanish if the contour is moved to
infinity and the contributions from $C_2$ and $C_5$ cancel each other. The
integrals along the semicircles around $-ib$ and the circle around $-i$ tend
to zero if the radius tends to zero. Due to the branch cut the contributions
from $C_3$ and $C_4$ are equal. Thus, the integral to be solved reads
\begin{eqnarray}
  \hspace{-6mm}I &\!=\!& \!\int_{-\infty}^\infty 
  \!\frac{e^{-ir'z}dz}{\sqrt{z^2+1}\sqrt{z^2+b^2}} = \int_{-ib}^{-i}\!
  \frac{ 2 e^{-ir'z}dz}{\sqrt{z^2+1}\sqrt{z^2+b^2}} 
\end{eqnarray}
and by changing the variable of integration ($z = -i-iy$), taking into account
that $b\gg 1$ in the Gross-Pitaevskii regime, we get
\begin{eqnarray}
  I & \approx & 2 e^{-r'} \int_{0}^{\infty} 
  \frac{e^{-yr'}dy}{\sqrt{2y+y^2}\sqrt{b^2-1-2y-y^2}} \; .
\end{eqnarray}
As we are looking for an approximation for $r'=\kt r \gg 1$, we notice that 
only small values of $y$ play an important role for the evaluation of the
integral. 

Hence we neglect the expression $1+2y+y^2$ in the second term in the
denominator which yields
\begin{eqnarray}
  I & \approx & 2 \frac{e^{-r'}}{\sqrt{b^2}}
  \int_{0}^{\infty} \frac{e^{-yr'}}{\sqrt{2y+y^2}}\,dy \;=\;
  \frac{2}{\sqrt{b^2}} \,K_0(r')
\end{eqnarray}
and as $\tilde{m}$ is an even function in $r$ we get the final result
\begin{eqnarray}
  \hspace{-5mm}\tilde{m}(r) & \approx & 
- \frac{\omm}{2\pi \kc} K_0(\kt|r|) \; \approx \; - \frac{\kc}{4\pi}  \;K_0(\kt|r|)\; .
\end{eqnarray}


\bibliographystyle{prsty}
\bibliography{References}

\end{document}